\documentclass[preprint]{aastex}
\usepackage{epsfig}

\slugcomment{Submitted to Astrophys. J.}
\shorttitle{Montaruli et al.}
\shortauthors{Neutrino astronomy in MACRO}

\begin{document}

\title{Neutrino astronomy with the MACRO detector}

\author{
\begin{center}
{\bf The MACRO Collaboration}\\
\end{center}
\nobreak\bigskip\nobreak
M.~Ambrosio$^{12}$, 
R.~Antolini$^{7}$, 
G.~Auriemma$^{14,a}$, 
D.~Bakari$^{2,17}$,
A.~Baldini$^{13}$, 
G.~C.~Barbarino$^{12}$, 
B.~C.~Barish$^{4}$, 
G.~Battistoni$^{6,b}$, 
R.~Bellotti$^{1}$, 
C.~Bemporad$^{13}$
% ...
\\
\vspace{1.5 cm}
\footnotesize
1. Dipartimento di Fisica dell'Universit\`a di Bari and INFN, 70126 
Bari,  Italy \\
2. Dipartimento di Fisica dell'Universit\`a di Bologna and INFN, 
 40126 Bologna, Italy \\
3. Physics Department, Boston University, Boston, MA 02215, 
USA \\
4. California Institute of Technology, Pasadena, CA 91125, 
USA \\
5. Department of Physics, Drexel University, Philadelphia, 
PA 19104, USA \\
6. Laboratori Nazionali di Frascati dell'INFN, 00044 Frascati (Roma), 
Italy \\
7. Laboratori Nazionali del Gran Sasso dell'INFN, 67010 Assergi 
(L'Aquila),  Italy \\
8. Depts. of Physics and of Astronomy, Indiana University, 
Bloomington, IN 47405, USA \\
9. Dipartimento di Fisica dell'Universit\`a dell'Aquila  and INFN, 
 67100 L'Aquila,  Italy \\
10. Dipartimento di Fisica dell'Universit\`a di Lecce and INFN, 
 73100 Lecce,  Italy \\
11. Department of Physics, University of Michigan, Ann Arbor, 
MI 48109, USA \\        
12. Dipartimento di Fisica dell'Universit\`a di Napoli and INFN, 
 80125 Napoli,  Italy \\        
13. Dipartimento di Fisica dell'Universit\`a di Pisa and INFN, 
56010 Pisa,  Italy \\   
14. Dipartimento di Fisica dell'Universit\`a di Roma ``La Sapienza" and INFN, 
 00185 Roma,   Italy \\         
15. Physics Department, Texas A\&M University, College Station, 
TX 77843, USA \\        
16. Dipartimento di Fisica Sperimentale dell'Universit\`a di Torino and INFN,
 10125 Torino,  Italy \\        
17. L.P.T.P., Faculty of Sciences, University Mohamed I, B.P. 524 Oujda, Morocco \\
18. Dipartimento di Fisica dell'Universit\`a di Roma Tre and INFN Sezione Roma Tre, 
 00146 Roma,   Italy \\ 
$a$ Also Universit\`a della Basilicata, 85100 Potenza,  Italy \\
$b$ Also INFN Milano, 20133 Milano, Italy\\
$c$ Also Istituto TESRE/CNR, 40129 Bologna, Italy \\
$d$ Also Scuola Normale Superiore di Pisa, 56010 Pisa, Italy\\
$e$ Also Universit\`a di Trieste and INFN, 34100 Trieste, 
Italy \\
$f$ Also Dipartimento di Energetica, Universit\`a di Roma, 
 00185 Roma,  Italy \\
$g$ Also Institute for Nuclear Research, Russian Academy
of Science, 117312 Moscow, Russia \\
$h$ Also Institute for Space Sciences, 76900 Bucharest, Romania \\
$i$ The Colorado College, Colorado Springs, CO 80903, USA\\
$m$ Also Department of Physics, Pennsylvania State University, 
University Park, PA 16801, USA\\
$n$ Also RPD, PINSTECH, P.O. Nilore, Islamabad, Pakistan\\
$o$ Also Dipartimento di Fisica dell'Universit\`a 
della Calabria, Rende (Cosenza), Italy \\
$p$ Also Department of Physics, James Madison University, Harrisonburg,
VA 22807, USA\\
$*$ Corresponding author, e-mail: montaruli@ba.infn.it
}

\begin{abstract}

High energy gamma ray astronomy is now a well established field and
several sources have been discovered in the region from a few GeV up 
to several TeV. If sources involving hadronic processes exist, the 
production of photons would be accompanied by neutrinos too.
Other possible neutrino sources could be related to the annihilation
of WIMPs at the center of galaxies with black holes.

We present the results of a search for point-like sources using 1100
upward-going muons produced by neutrino interactions in the rock below  
and inside the MACRO detector in the underground Gran Sasso Laboratory. 
These data show no evidence for a possible neutrino point-like
source or for possible correlations between gamma ray bursts and neutrinos.
They have been used to set flux upper limits for candidate point-like
sources which are in the range $10^{-14}-10^{-15}$
cm$^{-2}$ s$^{-1}$.
\end{abstract}


\section{Motivations for neutrino astronomy}
\label{sec:motiv}

The origin of cosmic rays is still largely an open question. 
The cosmic ray spectrum extends up to $10^{20}$ eV and the nature of the 
mechanisms capable of explaining such high energies is still unknown. 
Due to magnetic fields, charged cosmic particles are deflected from 
their original direction,
hence the information on the position of their source 
is lost.
On the other hand, protons of energies $\gtrsim 10^{7}$ TeV and 
neutral particles, such as photons and neutrinos, 
point back to their sources since they are not deflected by 
magnetic fields. 
However, the universe should become opaque to protons with energies above 
$\sim 5 \cdot 10^{19}$ eV at distances of $\sim 30$ Mpc 
due to photo-pion production when they interact with the Cosmic
Microwave Radiation (CMBR) \cite{GZK}.
Photons are currently the main observation channel of our universe
and the field of gamma-ray astronomy is now well established.

The idea of using neutrinos as probes of the deep universe
was introduced in the sixties. Already in those
years the first calculations on the diffuse neutrino flux from interactions
of cosmic rays in the Galaxy \cite{Greisen} 
and of high energy neutrinos from the Crab \cite{Bahcall64} 
were performed.

Neutrinos are weakly interacting particles and are therefore much 
less absorbed 
than gamma rays, which are not only absorbed during their propagation, but
can even be absorbed by the source producing them.
Neutrinos can bring information on the deep interior of sources and on the
far Universe. 
Several examples of detection of cosmic neutrinos already exist: 
solar neutrinos (from 0.1 MeV up to around 10 MeV), first detected
by Homestake \cite{Davis}, 
and neutrinos from SN1987A (from $\sim 10$ MeV up to 
$\sim 50$ MeV) detected by Kamiokande and IMB \cite{Hirata87,Bionta87}. 
Nevertheless, neutrinos of astrophysical origin with energies larger
than 100 MeV have not yet been observed. This observation would open
the new field of high energy 
neutrino astronomy complementary to gamma ray astronomy.
Moreover, an important hint on the existence of neutrino astronomy  
would come from the detection of photons of
energies above 100 TeV. Such energies, in fact, cannot be explained
by electron energy loss mechanisms (e.g. synchrotron radiation, bremsstrahlung
and inverse Compton scattering) because electron acceleration 
is limited by the intense synchrotron radiation produced in the ambient 
magnetic fields. Therefore, alternate acceleration mechanisms which involve 
neutrinos are required.

Satellites, ground based imaging Cherenkov 
telescopes and extensive air shower arrays are currently investigating
the universe, cosmic ray sources and acceleration mechanisms
using photons as probes \cite{Jackson93}.
Current Space experiments typically work in the energy range up 
to about 30 GeV and ground-based terrestrial experiments have 
typical threshold energies of about 250 GeV. 
Ground based experiments have larger surfaces and longer time exposures
than space experiments, therefore they can observe higher energies 
where fluxes are low. 
Nevertheless, they are limited at low energies by the large background due to 
gamma rays produced in the electromagnetic cascades induced by cosmic ray 
interactions in the atmosphere.

The EGRET detector on board the Compton Gamma Ray Observatory (CGRO) 
satellite has thus far furnished the largest amount of 
information on sources up to $\sim 30$ GeV. The recent third EGRET catalogue  
\cite{EGRET}, covering the observations made from 1991 to 1995, 
contains 271 sources observed with energies greater
than 100 MeV. Between them, there are
5 pulsars, one probable radio galaxy (Cen A), 
66 high-confidence identifications of blazars
(BL Lac objects and radio quasars), 27 lower-confidence potential blazar
identifications and a large number of identified supernova remnants (SNRs),
and also 170 sources not yet identified firmly with known objects. 

Satellite based detectors are providing observations
on $\gamma$-ray bursts (GRBs) capable of solving the mystery
concerning their nature.
The BATSE experiment \cite{BATSE} on CGRO satellite 
has now detected more than 2500 
$\gamma$-ray bursts and 
the Italian-Dutch {\it Beppo}SAX satellite \cite{Frontera} is
providing breakthroughs thanks to the precise measurement
(the error box radius is at the level of 4') of the position of the bursts.
Ground based experiments are looking for emissions above the TeV from
GRBs: recently the Milagrito detector has found a correlation
with the BATSE GRB970417a with chance probability 
$1.5 \times 10^{-3}$ \cite{Milagrito}.   

Cherenkov telescopes at ground level such as the Whipple observatory, HEGRA,
Cangaroo and University of Durham Mark 6 telescopes,
have so far detected 8 sources emitting $\gamma$-rays well above 
300 GeV: 
the supernova remnants Crab \cite{Lang}, Vela Pulsar (at a distance of only 
$\sim 500$ pc)\cite{Cangaroo0}, SN1006 \cite{Cangaroo}, 
the extra-galactic BL Lac objects (highly
variable active galactic nuclei) Mkn 421 ($z=0.031$)\cite{Punch,Hegra}, 
Mkn 501 ($z < 0.034$) \cite{Quinn}, PKS2155-304 ($z=0.116$) \cite{Chadwick},
and the pulsars PSR1706-44 \cite{Kifune}, PSR1259-63 \cite{Sako}. 
The Whipple group detected the first source, the Crab supernova remnant 
\cite{Lang}.  
The Crab is considered now as a
standard candle for high energy gamma ray astronomy due to its gamma ray 
steady emission. 

The number of sources detected so far by ground based experiments
is much smaller than the number of sources detected by EGRET. 
One of the possible explanations is that high energy gamma rays are 
absorbed: TeV $\gamma$-rays suffer absorption through pair production
in intergalactic space on infrared light, 
PeV $\gamma$s on the CMBR and EeV 
$\gamma$s on radio-waves. This can even explain why the only BL Lac 
objects observed till now are also the nearest ones.

The discovery of TeV gamma rays emitted from the 8 sources quoted above
shows the possibility of production from $\pi^{o}$ decay 
and the possible existence of beam dump sources 
(see Sec.~\ref{sec:astro}) producing high energy neutrinos. 
Nevertheless, ``a few TeV'' are  
energies not high enough to exclude a synchrotron
radiation production mechanism. It could be completely excluded
only for sources of energies above 100 TeV,
but up to now, no source emitting at such energies has been discovered. There 
were some claims in the past, particularly about Cygnus X3, but they have 
not been confirmed \cite{Hillas95}.

A different kind of neutrino production from astrophysical sources has
been suggested by 
Gondolo and Silk \cite{Gondolo}. If cold dark matter
exists in the Galactic Center, it can be accreted by the black hole
which probably is there. The cold dark matter is redistributed by the 
black hole into a cusp, which they call ``central spike''.
If dark matter is made of neutral particles that can annihilate, 
such as the supersymmetric neutralino, the annihilation rate in the spike 
is strongly increased as it depends
on the square of the matter density. Neutrinos can escape and produce
relevant fluxes. For the neutralino, the fluxes are very high
in the case of the presence of a central spike at the level of 
$10^{-15} \div 10^{-14}$ cm$^{-2}$ s$^{-1}$ for $m_{\chi} \gtrsim 50$
GeV.

During the '70s and '80s 
the first generation underground detectors  of surface 
$\sim 100$ m$^{2}$ have been measuring neutrinos. Those detectors
were aimed at detecting proton decay for which atmospheric neutrinos
were considered a background. Nevertheless, neutrinos were soon 
considered as an interesting signal themselves and results on 
searches for astrophysical sources of neutrinos were made.
Previous results on the search for point-like
sources have been published by the Kolar Gold Field experiment \cite{Adarkar}
and by the water Cherenkov detectors IMB \cite{IMB}
and Kamiokande \cite{Kamiokande}.
Other experiments (Baksan \cite{Baksan}
and AMANDA \cite{AMANDA}) have presented preliminary
results at conferences.
MACRO has been detecting muon neutrinos since 1989 while it was still 
under construction.
We present here the results of the search for astrophysical neutrino sources
with MACRO during the period March 1989 - September 1999.
\section{ Neutrino astronomy}
\label{sec:astro}

Astrophysical neutrinos can be produced in the interactions of protons 
accelerated by compact sources with a target around the source (gas of matter
or photons). This is the most plausible model for a neutrino source, the
so called ``beam dump model'' \cite{Berezinsky85,Gaisser95}.
The acceleration process requires the presence of a strong 
magnetic field with sufficient local gas to act as a beam dump.
The column density of the gas in the source
is assumed to be larger than the
nuclear depth ($x_{N} \sim 70$ g/cm$^{2}$), but smaller
than the neutrino absorption depth due mainly to $\nu N$ interactions
($x_{\nu} \sim 3 \cdot 10^{12} \cdot 100$ GeV$/E_{\nu}$  g/cm$^{2}$
and $x_{\bar{\nu}} \sim 6 \cdot 10^{12} \cdot 100$ GeV$/E_{\bar{\nu}}$
g/cm$^{2}$ \cite{Berezinsky85}).
The chain of reactions is:
\begin{eqnarray}
\begin{array}{cc}
p + N (\gamma)\rightarrow \pi^{0} \hspace{.5cm}+ \hspace{.5cm}
\pi^{\pm} + ... & \\
&\hspace{-7.cm}\searrow \gamma + \gamma \hspace{.5cm}  
\searrow \mu^{\pm} + \nu_{\mu}(\bar{\nu}_{\mu}) \\
&\hspace{-.6cm}\searrow e^{+} + \nu_{e} + \bar{\nu}_{\mu}  
\; \; (e^{-} + \bar{\nu}_{e} + \nu_{\mu}) \, .
\end{array}
\label{eq:source}
\end{eqnarray} 
Neutral pions produce the observed photons;
from the same chain it is expected the production of
charged pions and kaons which can decay producing neutrinos and muons. 
Moreover, muons decay too. 
The result are neutrinos and antineutrinos of electron  
and muon flavors.
Neglecting the photon absorption effect, which is subject to
very large uncertainties, the neutrino flux have at least the same
spectral shape and intensity with respect
to the gamma ray flux; hence very low neutrino event rates are expected
due to the small neutrino cross section. 
The presence of $\gtrsim 100$ TeV gamma ray sources should guarantee 
the existence of neutrino sources, 
but no reliable information could be drawn on neutrino
fluxes from gamma ray ones because they are
subject to non negligible absorption.

Cosmic accelerators produce a power law spectrum:
\begin{equation}
\frac{d\phi}{dE} \propto E^{-(\gamma+1)}
\end{equation}
where $\gamma \sim 1 + \varepsilon$, with $\varepsilon$ a small number.
The first order Fermi acceleration mechanism 
in strong shock waves has the attractive
feature of resulting in this kind of power spectrum \cite{Longair} and
it predicts a spectral index $\gamma \sim 1$. 
The primary cosmic ray spectrum is thought to be steeper 
than the one resulting from a cosmic accelerator because of the energy
dependence of the cosmic ray diffusion out of the Galaxy, as explained in  
\cite{Gaisser95}.  

Primary cosmic rays interact with the nuclei in the atmosphere and produce 
cascades from which atmospheric neutrinos of muon and electron flavor 
originate from the decays of pions, kaons and muons.
Up to now, only atmospheric neutrinos with energies above 100 MeV
have been detected by underground detectors.
If all the parent mesons of atmospheric neutrinos decay 
the neutrino spectrum follows the spectrum of the parent particles
(($\frac{d\phi_{\nu}}{dE})_{atm} \propto E^{-2.7}$ for $E_{\nu} \lesssim 10$ 
GeV). For higher energies,  
since the path length in the atmosphere is not large enough to allow the decay 
of all pions and kaons, interactions of mesons begin to dominate and
the atmospheric neutrino spectrum becomes steeper
(($\frac{d\phi_{\nu}}{dE})_{atm} \propto E^{-3.7}$ for $E_{\nu} \gtrsim 100$ 
GeV) due to the change of the spectral index of the meson spectra.
These neutrinos originating in the Earth atmosphere 
are a background for the search for astrophysical neutrinos 
which, on the other hand, are produced by cosmic rays at their acceleration
sites and hence should follow the hard cosmic ray source spectrum of the form:
\begin{equation}
\left(\frac{d\phi_{\nu}}{dE_{\nu}}\right)_{source}
\propto E^{-(2.0 \div 2.5)}  \,.
\label{eq:spectra}
\end{equation}
Thus the signal to noise ratio becomes larger at increasing energies, and above
some tens of TeV the neutrinos from sources start to dominate. 

The search presented here uses only the direction information of the neutrinos.
Other searches could maximize the signal to noise ratio
using the energy information on the detected particle 
and looking to the diffuse
neutrino events from the whole sky.
This was done by the Frejus experiment \cite{Frejus}
and some results have been recently
presented by the Baikal collaboration \cite{Baikal}. 
Preliminary MACRO results were presented elsewhere \cite{Corona95}
and will be the subject of a future paper.

\subsection{Candidate sources and expected rates}

High energy neutrinos are expected to be emitted from a wide class of
possible celestial objects which can be divided into two wide
classes: galactic sources and
extragalactic sources \cite{Gaisser95}.

Galactic sources are energetic systems, such as binary systems and supernova 
remnants, in which cosmic rays (CRs) 
are accelerated and interact with matter (mainly protons).
The most interesting sources are SNRs, which are the most likely
sources to be observed by a detector of the MACRO size. 
In such systems, the target 
is the material of the expanding shell and the accelerating mechanism is 
originated by the intense magnetic field of the pulsar.
There are however possibilities to have neutrino emissions originate
by acceleration at the supernova blast waves and therefore neutrino emission 
even without pulsars. 
The neutrino emission should be in an active time of up to a 
few years. Of course the disadvantage of galactic supernovae as neutrino 
emitters is that their rate is low (of the
order of 1/30 years). 
According to detailed calculations made for several historical supernova 
\cite{Gaisser96}, the most intense source should be the
supernova remnant Vela Pulsar with a rate of upward-going muons induced
by neutrinos in the rock surrounding a detector of the order 
0.1 ev/yr/1000 m$^{2}$ for $E_{\mu}>$1 GeV.
Another model for young SNRs with a pulsar having 
high magnetic field and
short period ($\sim 5$ ms) is suggested in Ref.~\cite{Protheroe98}:
for a beaming solid angle of neutrino emission of 1 sr, about 5 events/yr are
expected in 1000 m$^{2}$ for $E_{\nu} \ge 100$ GeV after 0.1 yr from an 
explosion at a distance of 10 kpc.

A different kind of galactic source is suggested in Ref.~\cite{Gondolo}
due to WIMP annihilations in the core
of the Galactic Center. The rates would be very promising,
even for detectors like MACRO, being of
the order of 1-20 events/yr/1000 m$^{2}$.

Possible extragalactic sources are active galactic nuclei and gamma-ray
bursters. For these sources the dominant mechanism for producing neutrinos 
is accelerated protons interacting on ambient photons.
Possible alternative mechanisms are
the so called Top-Down models \cite{Sigl}.
 
Active galactic nuclei (AGNs), being among the most luminous objects in the 
Universe with luminosities ranging from 10$^{42}$ to 10$^{48}$ erg/s,
have been recognized for a long time as promising possible sources
of neutrinos. Present models assume that they consist  
of a central engine (massive black hole) with an accretion disk and jets
\cite{Gaisser95}.
Accretion onto the central black hole provides the total power.
Two possible sources of high energy neutrino fluxes
within AGNs have been suggested. The first is associated with the
central engine and the second with the production in jets associated 
with several blazars (radio-loud AGNs in which the beam intersects
the observer line of sight).
AGNs could emit neutrinos up to $\sim 10^{10}$ GeV. 

Even considering the highest luminosities and the presence of jets, 
single AGNs are difficult to detect. Jets carry about 10$\%$ of the AGN 
luminosity and AGNs may appear brighter because of the motion of the
emitting matter toward the observer (for an observer looking along
the jet axis $E_{obs} = \Gamma E_{jet}$ and
$L_{obs} = \Gamma^{4} L_{jet}$, where $\Gamma$ is the Lorentz factor).
Expected event rates for blazars are of the order of
$10^{-2} - 10^{-1}$ /1000 m$^{2}$/yr for $\Gamma = 10-10^{2}$
and $E_{\nu}> 1$ TeV \cite{Halzen98}. 

Stecker {\sl et al.} \cite{Stecker} suggested the possibility to integrate the
neutrino flux from single generic AGNs to obtain a diffuse flux from all
cosmological AGNs.
Various models have been suggested \cite{Stecker,Szabo} \footnote{Most 
of the Szabo and Protheroe \cite{Szabo} models are excluded by the Frejus
limit \cite{Frejus}.} 
and the event rates in upward-going muons 
vary between $\sim 10^{-1} - 10$ /1000 m$^{2}$/yr for $E_{\nu}> 1$ TeV.

Gamma Ray Bursters 
are considered as promising sources of high energy neutrinos.
They yield transient events originating beyond the solar system, with typical
durations of $10^{-2}\div 10^{3}$ s. The BATSE \cite{BATSE}
experiment has now collected more than 2500 events which appear
isotropically distributed. This feature suggests that they are
located at cosmological distances.
The recent observations by {\it Beppo}SAX of  GRB970228 have
allowed the precise measurement of the position which for the
first time led to the identification of a fading optical counterpart 
\cite{Costa}. Immediately after, the direct measurement of the redshift 
in the optical afterglow at $z=0.835$ for GRB970508 \cite{Metzger} and 
other identifications of the distances of GRBs have given support
to the cosmological origin hypothesis. These observations make GRBs the most 
luminous objects observed in our universe with emitted 
energies $\gtrsim 10^{51}$ erg and 
a spectrum peaked between 100 keV $\div$ 1 MeV.

One of the most plausible models is the ``fireball model'', which solves the
compactness problem introducing a beamed relativistic motion with 
$\Gamma \gtrsim 100$ of an expanding fireball \cite{Piran}. 

The question of the energy of the engine of GRBs, which is strictly
connected to that of beaming, is still under discussion. Evidence for
beaming are a break and a 
steepening of the spectrum.
They have been found in spectra of some 
bursts, e.g. GRB980519 and GRB990123; 
in the case of GRB990123 at $z=1.6$ for isotropic emission the emitted 
energy would be the highest ever observed ($2 \cdot 10^{54}$ erg)
while if there is a beam the emitted energy would be reduced 
to $\sim 10^{52}$ erg due to the Lorentz factor.   

Several authors have suggested a
possible correlation between Gamma Ray Bursters  
and emissions of high energy neutrinos \cite{Halzen,Bahcall,Meszaros,Vietri} 
produced by accelerated protons on photons. 
Expected rates could be up to $10^{6}$ muon induced events in a 
1000 m$^{2}$ detector for muon energies above $\sim$ 30 TeV for emissions
lasting $< 1$ s \cite{Halzen}. 
In other scenarios, such as for fireballs, rates
are of the order of $\sim 10^{-3}$ upward-going muons in a 
1000 m$^{2}$-size detector for a burst at a distance of 
100 Mpc producing $0.4 \cdot 10^{51}$ erg in $10^{14}$ eV neutrinos \cite{WB}.
Considering a rate of $10^{3}$ bursts per year over 4$\pi$ sr, 
averaging over burst distances and energies, $\sim 2 \cdot 10^{-2}$ 
upward-going muons are expected in 1000 m$^{2}$ per 1 yr for 4$\pi$ sr 
\cite{Bahcall}. 
It is important to consider that the uncertainty on the Lorentz factor $\Gamma$
produces high variations in the expected rates: the higher the $\Gamma$, the
larger the luminosity at the observer ($L_{obs} \sim \Gamma^{4} L_{jet}$),
but the smaller are the rates of events because the actual photon 
target density in the fireball is diluted by 
large Lorentz factors (the fraction of total energy going into pion production
in the source and hence into neutrinos
varies approximately as $\Gamma^{-4}$ \cite{Halzen99}).
 
According to Waxman and Bahcall (WB) \cite{WB} an
energy independent upper bound on diffuse fluxes of neutrinos 
with $E_{\nu} \gtrsim 10^{14}$ eV produced 
by photo-meson or p-p interactions in sources from which protons can escape
can be estimated at the level of
$E_{\nu}^{2}\phi_{\nu} < 2 \times 10^{-8}$ GeV cm$^{-2}$ s$^{-1}$ sr$^{-1}$. 
This bound relies on the flux measurement of extremely 
high energy cosmic rays in extensive air showers, which are assumed to be
of extragalactic origin. Their limit would
exclude most of the present models of neutrino production in AGNs 
which are commonly
normalized to the extragalactic MeV-GeV gamma-ray background.
Contrary to WB, Mannheim, Protheroe and Rachen \cite{MPR}  
find an energy dependent upper limit which agrees within a factor of 2
with WB in the limited range of $E_{\nu} \sim 10^{16-18}$ eV,
while at other energies the neutrino flux is mainly limited by their
contributions to extragalactic gamma-ray background which is at a level of
about 2 orders of magnitude higher than the WB limit. 

\section{The MACRO detector and the data selection}

In the range of energies from several GeV to several TeV 
the neutrinos produced by astrophysical sources can be detected
in underground detectors as upward-going muons produced by 
neutrino charged current (CC) 
interactions in the rock surrounding the detector.
Neutrino events can be discriminated from among the background of atmospheric
muons of many orders of magnitude larger 
($\sim 5 \cdot 10^{5}$ at MACRO depth) 
recognizing that they travel from the
bottom to the top of the apparatus after having transversed the Earth.
Neutrino detection is experimentally much more difficult 
than the gamma ray one; because of the low neutrino interaction cross
section it requires very large detectors.

The MACRO detector, shown in Fig.~\ref{fig1} and 
described in detail in \cite{Ahlen93}
is located in the Hall B of the Gran Sasso underground laboratory at 
a minimum rock depth of 3150 hg/cm$^{2}$ and an average rock depth of 3700 
hg/cm$^{2}$.
The detector, 76.6 m long, 12 m wide and 9.3 m high,
is divided longitudinally in six similar supermodules and
vertically in a lower part (4.8 m high) and an upper part
(4.5 m high).

The active detectors include 14 horizontal and 12 vertical planes 
of 3 cm wide limited streamer tubes for particle tracking, 
and liquid scintillation counters for fast timing.
In the lower part, the eight inner planes of limited streamer tubes
are separated by passive absorbers (iron and rock $\sim 50$ g cm$^{-2}$)
in order to set a minimum threshold of $\sim 1$ GeV for vertical muons
crossing the detector. The upper part of the detector
is an open volume containing electronics and other equipment.
The horizontal streamer tube planes are instrumented with
external 3 cm pick-up strips at an angle of $26.5^{\circ}$ with respect 
to the streamer tube wires, providing stereo readout of the detector hits.
The transit time of particles through the detector is measured by the
time of flight technique (T.o.F.) using
scintillation counters. The mean time at which 
signals are observed at the two ends of each counter is measured 
and the difference in the measured mean time between counters 
located in different planes gives the T.o.F.. 
The time resolution of the scintillation counter system
is about 500 ps.

In order to achieve the largest reconstruction 
efficiency for all directions, three
algorithms for muon tracking are used in this analysis. 
The first kind of tracking is for events with aligned hits
in at least 4 horizontal planes; 
the second is for events with at least 2 horizontal planes in
coincidence with at least 3 vertical planes;
the third is for events having at least 3 vertical
planes in coincidence with two scintillation counters (this tracking
is useful for almost horizontal tracks). 

The angular resolution depends on the wire and strip 
cluster widths and on the track length. 
The average errors on the slopes of tracks are 
$0.14^{\circ}$ for the wires and $0.29^{\circ}$ for the strips \cite{Ahlen93}. 
Our pointing capabilities for point-sources 
has been checked with the observation of the Moon shadowing effect
using atmospheric down-going muons \cite{Moon98}. 

The data used for the upward-going muon search belong to three running
periods with different apparatus configurations: 26 events
have been detected with the lower half of the
first supermodule from March 1989 until November 1991 (about 1/12 of 
the full acceptance, livetime of
1.38 years, efficiencies included), 55 with the full lower half of 
the detector (about 60$\%$ of the full acceptance) from December 1992 
until June 1993 (0.41 years, efficiencies included).
Starting from April 1994 the apparatus has been running in the final
configuration.
From April 1994 until Sep. 1999 we have measured 1000 events with 
the full detector (4.41 live years, including efficiencies).
We also consider events which were measured during periods 
when the detector acceptance was changing with 
time due to construction works (19 events during 1992, 0.2 yr).

The selection of upward-going muons 
using the T.o.F. technique has been described in detail in
Ref.~\cite{Ambrosio95,Ambrosio98}.
The velocity and direction of muons is determined 
from the T.o.F. between at least
2 scintillation layers combined with the path length
of a track reconstructed using the streamers. Taking as a reference the
upper counter which measures the time $T_{1}$, the time of flight
$\Delta T = T_{1} - T_{2}$ is positive if the muon travels downward
und it is negative if it travels upwards. 
Fig.~\ref{fig2} shows the $1/\beta = c \Delta T/L$ ($L$ is the track-length,
and $c$ the speed of light) distribution for the entire
data set. In this convention,  
muons going down through the detector have $1/\beta 
\sim 1$, while muons going upwards have $1/\beta \sim -1$.
Several cuts are imposed to remove backgrounds caused by
radioactivity in coincidence with muons and multiple muons.
The main cut requires that the position 
of a muon crossing a scintillator agrees within 70 cm
(140 cm for slanted tracks with $\cos\theta \le 0.2$) 
with the position along the
counter determined by the more precise streamer system.  
Other cuts apply only to events which cross 2 scintillator planes only.
These cuts tend to remove high multiplicity events because when
more than one track crosses the same scintillator box the reconstructed time 
of the event is wrong. 
Events which cross more than 2 scintillator planes (about 50$\%$ of the
total) have a more reliable time determination thanks to the possibility
to evaluate the velocity of the particle from
a linear fit of times as a function of the height of the scintillator
counters.
In this case, the only cut then is on the quality of the fit 
($\chi^2 \le 10$).

Events in the range $-1.25 < 1/\beta < -0.75$ 
are defined to be upward-going muon events.
There are 1100 events which satisfy this definition summed over all 
running periods.
One event is shown in Fig.~\ref{fig3}.
In order to maximize the acceptance for this search, we do not
require a minimum amount of material be crossed by the muon track
as was done to select the sample used for the neutrino oscillation
analysis \cite{Ambrosio95,Ambrosio98}.
Without this requirement we introduce some background
due to large angle pions produced by down-going muons \cite{Spurio98}.
We also include events with an interaction vertex inside the lower
half of the detector.
All of these data can be used for the point-like astrophysical source search
since the benefit of a greater exposure 
for setting flux limits offsets the slight
increase of the background and of the systematic error in the acceptance.
Moreover, one can notice that for neutrino oscillation studies 
upward-going muons are mostly signal and background rejection is very 
critical, while for neutrino astronomy upward-going 
muons are mostly background due to atmospheric neutrinos and background 
rejection is less critical.
 
\section{Neutrino signal in upward-going muons}

Muon neutrinos are detected as upward-going muons through CC interactions:
\begin{equation}
\nu_{\mu} (\bar{\nu}_{\mu}) + N \rightarrow \mu^{-} (\mu^{+}) + X \,.
\end{equation}
The probability that a neutrino (or antineutrino) 
with energy $E_{\nu}$ interacts in the rock below the detector and gives rise
to a muon which crosses the apparatus with energy 
$E_{\mu} \ge E_{\mu}^{th}$ ($E_{\mu}^{th}$ is the energy 
threshold of the apparatus) is:
\begin{equation}
P_{\nu}(E_{\nu},E^{\mu}_{th}) = N_{A} \int_{0}^{E_{\nu}} \, dE^{'}_{\mu} \,
\frac{d\sigma_{\nu}}{dE^{'}_{\mu}}(E^{'}_{\mu},E_{\nu}) \cdot 
R_{eff}(E^{'}_{\mu},E^{\mu}_{th})
\label{eq:p}
\end{equation}
where $N_{A}$ is Avogadro's number. 
This probability is a convolution of the $\nu$ cross sections and
of the muon effective range $R_{eff}(E_{\mu}, E_{\mu}^{th})$ described below; 
the computed probability is shown in Fig.~\ref{fig4}
and some values are given in Tab.~\ref{tab1}.
The trend of the probability at energies $\lesssim$ 1 TeV reflects 
the cross-section linear rise with neutrino energy ($\sigma_{\nu} 
\propto E_{\nu}$) and that of the muon range ($R_{eff} \propto E_{\mu}$),
while at higher energies it reflects the damping effect of the propagator 
($\sigma_{\nu} \propto E_{\nu}^{0.4}$ for $E_{\nu} \gtrsim 10^{3}$ TeV)
and the logarithmic rise of the muon range with its energy.

It is relevant to notice that, thanks to the recent HERA measurements
\cite{Wolf}, our knowledge of the high energy deep inelastic neutrino
cross section has improved significantly.
There is good agreement between various sets 
of parton functions which provide confident predictions of the
cross-sections up to $10^{6}$ GeV \cite{Gandhi}. 
For the calculation of the probability shown in Fig.~\ref{fig4}
we have used the CTEQ3-DIS \cite{CTEQ} parton function, available in the
PDFLIB CERN library \cite{PDFLIB}, which have been 
considered by Gandhi {\sl et al.}  \cite{Gandhi} and in good agreement with the
more recent CTEQ4-DIS.

The technique of detecting upward-going muons generated in the
rock surrounding a detector has the advantage to increase
the effective detector mass, which in fact is a convolution
of the detector area and of the muon range in the rock.
The gain increases with energy: for example, for 
TeV muons, the range is of the order of 1 km. 
The effective muon range is given by the probability that
a muon with energy $E_{\mu}$ survives with energy above threshold
after propagating a distance X:
\begin{equation}
R_{eff}(E_{\mu},E^{\mu}_{th}) = \int_{0}^{\infty} dX \, P_{surv}(E^{'}_{\mu},
E^{th}_{\mu},X)
\end{equation}
where the integral is evaluated from the $\mu$ energy losses. 
We have used the energy loss 
calculation by Lohmann {\it et al.} \cite{Lohmann85} using standard rock
for muon energies up to $10^5$ GeV. For higher energies we use the
approximate formula:  
\begin{equation}
\frac{dE_{\mu}}{dX} = \alpha + \beta \cdot E_{\mu} \, ,
\end{equation}
where $\alpha \sim 2.0$ MeV g$^{-1}$ cm$^{2}$ 
takes into account the continuous 
ionization losses and $\beta \sim 3.9 \cdot 10^{-6}$ g$^{-1}$ cm$^{2}$ takes
into account the 
stochastic losses due to bremsstrahlung, pair production and
nuclear interactions.

The flux of neutrino induced muons detected by an apparatus for a source of 
declination $\delta$ and for
a neutrino spectrum $\Phi_{\nu}(E_{\nu}) \propto E^{-\gamma}$ is:
\begin{eqnarray}
\Phi_{\mu}(E^{th}_{\mu},E_{\nu},\delta)=  N_{A}
\int_{E^{th}_{\mu}}^{E^{max}_{\mu}}
\frac{d\sigma_{\nu}}{dE^{'}_{\mu}}(E^{'}_{\mu},E_{\nu}) \cdot
R_{eff}(E^{'}_{\mu},E^{\mu}_{th}) \cdot Area(E^{'}_{\mu},\delta) 
\cdot \Phi_{\nu}(E_{\nu}) \; dE^{'}_{\mu} \, .
\label{eq:resp}
\end{eqnarray}
The effective area of the detector $Area(E^{'}_{\mu},\delta)$, averaged over
24 hours, depends on the source declination. In the low energy region,
the effective area increases with increasing muon energy because not all 
muons are detected depending on their track length in the detector. At higher
energies (in MACRO for $E_{\mu} \gtrsim 3$ GeV)
it reaches a plateau when all muons from all directions have enough energy 
to be detected. At very high energies the effective area can decrease 
due to electromagnetic showers. As a matter of fact,
the efficiency of the analysis cuts can decrease due to high track 
multiplicities for high energy events. Moreover the presence of showers
could lead to a bad reconstruction of the neutrino induced muon 
with another track of the shower.
From Monte Carlo studies, the MACRO average effective area 
begins to decrease for $E_{\mu} \gtrsim 1$ TeV and it is 
about 20$\%$ (42$\%$) lower at 10 TeV (100 TeV) with respect to 10 GeV. The
average area as a function of declination for various energies
is shown in Fig.~\ref{fig5}. It has been obtained using the detector
simulation based on GEANT \cite{Brun87}, but
modified to properly treat the stochastic
muon energy losses above 10 TeV \cite{Perrone}.
To obtain large Monte Carlo statistics we have used ``beams'' 
of monoenergetic muons intercepting isotropically from the lower hemisphere
a volume containing MACRO and more than 2 m of the surrounding 
rock (to evaluate the effect of electromagnetic showers induced by 
high energy muons). For each beam energy, we have simulated about 10$^{5}$
muons.

Due to the increasing value of the $\nu$ cross-section, at high energies
neutrinos are ``absorbed'' by the large amount of material they cross through 
the Earth.
Neutrino absorption in the Earth can be taken into 
account introducing in the integral
in eq.~\ref{eq:resp} the exponential factor:
\begin{equation}
e^{-N_{A} \cdot \sigma_{\nu}(E) \cdot X(\cos\theta)}
\end{equation}
which depends on $X(\cos\theta)$, the quantity of matter transversed by
the incident neutrino in the Earth and hence on its zenith angle. 
The differential number of neutrinos as a function of the
neutrino energy (response curves) for a source of differential 
spectral index $\gamma=2.1$ 
at two different declinations 
with and without absorption in the Earth is shown
in Fig.~\ref{fig6}. 
The median neutrino energy 
is about 15 TeV, while for the atmospheric neutrinos
it is between 50-100 GeV \cite{Ambrosio98}.
It is noticeable how the
absorption becomes negligible for sources
seen near the horizon ($\delta \sim 0^{\circ}$). 
In Fig.~\ref{fig7} the same response curves
are shown for 3 spectral indices. 
In these plots, the normalization of the neutrino fluxes is arbitrary.
 
It is relevant to notice that if muon neutrinos oscillate
into tau neutrinos, as atmospheric neutrino
experiment results suggest \cite{Ambrosio98,SK}, 
$\nu_{\tau}$ are subject to considerably
less absorption than muon neutrinos \cite{Bottai}. Tau neutrinos
are subject to a regeneration effect in the Earth:
$\nu_{\tau}$ interacts and the produced tau lepton 
immediately decays with negligible energy loss; hence from $\tau$
decay another $\nu_{\tau}$ is produced. 
This effect, more noticeable for harder spectra,
has been neglected here.

The fluxes of detectable upward-going muons for sources with $\gamma=2.1$
and $\delta = -60^{\circ}, 0^{\circ}$ are shown with and without absorption
in Fig.~\ref{fig8}.
If one assumes that the normalization of the neutrino flux is
of the order of the upper limits from $\gamma$-ray experiments 
at $\sim 100$ TeV ($2 \times 10^{-13}$ cm$^{-2}$ s$^{-1}$
for the Galactic Center), 
the expected rate of neutrino induced muons varies between 
$10^{-2}\div 10^{-3}$ ($10^{-1}\div 10^{-2}$) 
events/yr/1000 m$^{2}$ for $\gamma=2.1$ ($2.5$) 
depending on the declination of the source.
Note that the softer the source spectrum, the higher the
neutrino event rates.

An important quantity in the search for celestial point sources is the 
effective angular spread of the detected muons with respect to the neutrino 
direction.
We have computed the angle between the neutrino and the detected 
muon using a Monte Carlo simulation.  We have assumed 
a neutrino energy flux of the form $dN/dE_{\nu} = 
constant \times E^{-\gamma}$, for several neutrino spectral indices $\gamma$,
and considered the neutrino cross-sections, 
the muon energy loss in the rock and
the detector angular resolution.
Tab.~\ref{tab2} shows the fraction of the events in a $3^{\circ}$ search 
half-cone for two different spectral indices as a function of the zenith angle.
With the simulations of monoenergetic muon beams on a box larger than 
the detector including $\sim 2$ m of rock, we have calculated the effective
area and 
even checked that our intrinsic resolution does not worsen with energy
due to the effect of increasing electromagnetic showers induced by 
stochastic energy losses of muons. Up to 100 TeV the average angle
between the generated muons and the reconstructed ones is less than 
$1^{\circ}$.

\section{Search for point-like sources}

The MACRO data sample is shown in equatorial coordinates
(right ascension in hours and declination in degrees) in Fig.~\ref{fig9}.
For the point-like source search using the
direction information of upward-going muons,
we evaluate the background due to atmospheric neutrino induced muons 
randomly mixing for 100 times the local angles of upward-going events
with their times. The number of mixings is chosen to have a statistical
error for the background about 10 times smaller than the
data fluctuations. The local angles are then smeared by $\pm 10^{\circ}$ 
in order to avoid repetitions, particularly in the declination regions
where there is small acceptance. The value of $10^{\circ}$ is chosen to
have variations larger than the dimensions of the search cones.

For a known candidate point-like source $S$
the background in the search cone $\Delta \Omega = \pi \omega^{2}$,
with $\omega$ the half width of the search cone in radians, 
is evaluated counting the events in a declination band around the 
source declination $\delta_{S}$ of $\Delta \delta = \pm 5^{\circ}$:
\begin{equation}
N_{back} = \frac{N(\Delta \delta) \Delta \Omega}{2\pi [\sin(\delta_{S}
+5^{\circ})-\sin(\delta_{S}-5^{\circ})]} \, .
\end{equation}

We have considered the case of a possible detection of an
unknown source represented by an excess of events clustered inside
cones of half widths $1.5^{\circ}$, $3^{\circ}$ and $5^{\circ}$.
Hence we have looked at the number of events falling inside these cones
around the direction of each of the 1100 measured events.
The cumulative result of this search is shown in Fig.~\ref{fig10} for
the data (full circles) and the simulation of atmospheric events (solid line).
We find 60 clusters of $\ge 4$ muons around a given muon (including
the event itself), to be compared with 56.3 expected from the 
background of atmospheric neutrino-induced muons.
The largest cluster is made of 7 events in the $3^{\circ}$ half-cone
and it is located around the equatorial coordinates 
(right ascension, declination)=($222.5^{\circ},-72.7^{\circ}$).
Other 2 clusters of 6 events in $3^{\circ}$ are located around
= ($188.1^{\circ},-48.1^{\circ}$) and ($342.5^{\circ},-74.4^{\circ}$), 
respectively.
Nevertheless, they are not statistically significant.

For our search among known point-sources, we have considered
several existing catalogues: the recent EGRET catalogue \cite{EGRET}, 
a catalogue of BL Lacertae objects
\cite{Padovani} of which 181 fall in the visible sky of MACRO ($-90^{\circ}
\le \delta \lesssim 50^{\circ}$), the list of 8 sources in the visible
sky emitting photons above TeV already mentioned in Sec.~\ref{sec:motiv}, 
the Green catalogue \cite{Green} of SNRs, the BATSE \cite{BATSE} catalogues, 
32 {\it Beppo}SAX GRBs \cite{Pian}, 
a compilation of 29 Novae X \cite{Masetti}, 
which are binaries with a compact object and a companion star which
transfers mass into an accretion disk. Novae X are characterized by sudden
increases of luminosities in the X range ($L \sim 10^{37}-10^{38}$ erg
s$^{-1}$ reached after 20-90 days).
Among these catalogues we have selected 42 sources we consider
interesting because they have 
the features required by the ``beam dump'' model. 
In Fig.~\ref{fig11} the distribution of the numbers of
events falling in the search cones is shown for the data
and the simulation for the 42 sources.
We find no statistically significant excess from any of the considered
sources with respect to the atmospheric
neutrino background. For the 42 selected sources
we find 11 sources with $\ge 2$ events in a search cone of
$3^{\circ}$ to be compared to 12.0 sources expected from the 
simulation. 
 
Upper limits on muon fluxes from sources can be calculated at a given 
confidence level, e.g. 90$\%$ c.l. as:
\begin{equation}
\Phi(90\% c.l.) = \frac{{\rm Upper \; limit} 
(90\% c.l.)}{{\rm effective \, area} \times {\rm livetime}} 
\, ,
\label{eq:limit}
\end{equation}
where the numerator is the upper limit calculated from the number of measured
events and from the number of expected background events and 
the denominator is the exposure of the detector which 
is the area of the apparatus seen by the source
during the livetime. 
Different methods to evaluate upper limits are described in 
\cite{DPB}.
We have calculated the upper limits (the numerator in eq.~\ref{eq:limit})
using the recent and well motivated unified approach by Feldman and Cousins
\cite{Feldman98}. It is possible to calculate neutrino
flux upper limits from muon flux upper limits  
because they are related (see eq.~\ref{eq:resp}).
The $90\%$ c.l. muon and neutrino flux limits are given in Tab.~\ref{tab3}
for the 42 selected sources. These limits are valid for muon
energies $> 1$ GeV. They include the effect of the absorption
of muon neutrino in their propagation through the Earth.
The limits are obtained assuming a neutrino
spectrum from a source with $\gamma = 2.1$.
Moreover, the effect of the decrease in efficiency at very high energies
and the reduction factors for a search half-cone of $3^{\circ}$ and a 
spectral index $\gamma = 2.1$ (given in Tab.~\ref{tab2}) are included. 
For comparison we include the best limits from previous experiments. 
In order to see how the limits depend on the spectral index 
$\gamma$ we report in Tab.~\ref{tab4} the percentage difference of the 
exposure as a function of declination calculated for a source with
$\gamma = 2.1$ and a source with $\gamma = 2.3$, 2.5, 2,7, 3.7.

To evaluate the physical implications of our limits we recall
that a muon flux of the order of $0.03 \times 10^{-14}$ cm$^{-2}$ s$^{-1}$
is expected from the supernova remnant Vela Pulsar \cite{Gaisser96}, which
predict a yield of neutrinos at the level of about one order of magnitude 
lower than present limits.

We notice that there are 6 events from GX339-4 in a $3^{\circ}$ with chance
probability $P= 6 \cdot 10^{-3}$. Considering that
we have looked at 42 sources the probability to find
such an excess from at least one of these sources is $8.6\%$
(evaluated from Fig.~\ref{fig11}).

Between the selected 42 candidate sources, Mkn 421 and Mkn 501 are 
particularly interesting due to the strong emissions (in the TeV region) 
they present. These emissions have variable intensity during time.
Mkn 421 shows a strongly variable emission with peak flares during
June 1995, May 1996 and April 1998 \cite{Krennrich99}. 
Mkn 501 had a high state emission during about 6 months in 1997, particularly
intense between April and September \cite{Protheroe97}. The strongest flare
in 1998 occurred on March 5.
Unfortunately, the MACRO esposure for this sources is not favoured
because they are seen almost at the horizon where the acceptance is
lower.
No event from both sources is found inside a search cone as large as 
$5^{\circ}$. Only two events for each of the sources are found inside
$10^{\circ}$. They are of marginal interest due to the large angle with
respect to the source directions. They have been measured in
periods in which there were not known intense flares (for Mkn 421:
10 Sep. 1996 and 27 Jun. 1998; for Mkn 501: 29 Sep. 1996 and 26 Jun. 1998).

We have also made a search for neutrino signals using a cumulative analysis:
for each of several catalogues of source types, we set a limit on flux
from sources from that catalogue.
In some situations (for example for a uniform distribution in space
of sources having the same intensity) this method could give
a better sensitivity than the search for a single source. It depends
on the spatial distribution and on the intensity of the sources.

We consider the average value $N_{0}$ of the distribution for the data 
and the average value $N_{B}$ of the distribution for the simulation in 
Fig.~\ref{fig11} for the 42 sources in the MACRO list, in Fig.~\ref{fig12} 
for the 220 SNRs,
in Fig.~\ref{fig13} for the 181 blazars and in analogous plots for the other 
catalogues.
Then we estimate the cumulative upper limits for $N$ sources in the 
catalogue as:
\begin{equation}
\Phi_{cumulative}(90\% {\rm c.l.}) = \frac{\rm Upper \; limit 
(90\% {\rm c.l.})}
{{\rm Average \; Area} \cdot {\rm livetime}}
\label{eq:newlim}
\end{equation}
where the average area is $\frac{\sum_{i=1}^{N} Area(\delta_{i})}{N}$,
with $Area(\delta_{i})$ the area seen by a source with 
declination $\delta_{i}$.
The upper limit is evaluated for $N_{0}>N_{B}$ as:
\begin{equation}
{\rm Upper \; limit} (90\% {\rm c.l.}) = 
N_{0} - N_{B} + 1.28 \cdot RMS/\sqrt{N} 
\label{eq:upp1}
\end{equation}
where RMS is the root mean square
value of the considered expected distributions and if $N_{0}<N_{B}$ as:
\begin{equation}
{\rm Upper \; limit} (90\% {\rm c.l.}) =  1.28 \cdot RMS/\sqrt{N}
\label{eq:upp2}
\end{equation}

We obtain $\Phi_{lim}(90\%) = 3.06 \cdot 10^{-16}$ cm$^{-2}$ s$^{-1}$
for the 42 sources in the MACRO list.
This can be considered a limit on a diffuse flux.  

In the case of the 220 SNRs in the Green catalogue 
we obtain the cumulative upper 
limit from the cumulative analysis shown in Fig.~\ref{fig12} of
$2.63 \cdot 10^{-16}$ cm$^{-2}$ s$^{-1}$.
This can be considered a diffuse muon flux limit for neutrino
production from supernova remnants.
For the 181 blazars in \cite{Padovani} (see Fig.~\ref{fig13})
we find a cumulative upper limit
of the muon flux $5.44 \cdot 10^{-16}$ cm$^{-2}$ s$^{-1}$.
In Tab.~\ref{tab5} we summarize the upper limits for 
the various catalogues considered.

Finally, it is interesting to note, that most of
the models for neutralino annihilation on the
Galactic Center in \cite{Gondolo} are excluded by our experimental
upper limit of $\sim 3 \cdot 10^{-15}$ cm$^{-2}$ s$^{-1}$ when there is
a ``central spike'' for a $3^{\circ}$ cone.
In Tab.~\ref{tab6} the muon flux limits (90$\%$ c.l.) for various search cones
around the direction of the Galactic Center ($3^{\circ}$, $5^{\circ}$
and $10^{\circ}$) for 5 values of the neutralino mass from 60 GeV
to 1 TeV are calculated. The dependence of the effective area of the detector
as a function of the neutrino energy has been calculated folding with the
neutrino flux from neutralino annihilation calculated by Bottino {\sl et al.} 
\cite{Bottino}. As a first approximation, the difference of using
the neutrino fluxes by Silk and Gondolo in the effective area calculation 
should be negligible.  
The limits are calculated for $E_{\mu}>1$ GeV.

\section{Search for correlations with gamma ray bursts}

We look for correlations with the gamma ray bursts given in the
BATSE Catalogues 3B and 4B \cite{BATSE} containing 2527 gamma ray
bursts from 21 Apr. 1991 to 5 Oct. 1999. They overlap in time with
1085 upward-going muons collected by MACRO during this period.
The effective 
area for upward-going muon detection in the direction of the bursts
averaged over all the bursts in the catalogue is 
121 m$^{2}$. Its value
is small because our detector is sensitive to neutrinos only in one hemisphere
and because it was not complete in the period 1991-1994. 
Fig.~\ref{fig14} shows our neutrino events 
and BATSE GRBs as a function of the year.

We find no statistically significant correlation between neutrino events
and gamma burst directions for search cones of 
$10^{\circ}$, $5^{\circ}$ 
and $3^{\circ}$ half widths. The width of the search cones is
related to the BATSE angular resolution; these cones include 
96.9$\%$, 85.1$\%$ and 70.5$\%$ neutrinos respectively if emitted from GRBs.
These numbers do not include the contributions due to 
the muon-neutrino angle,
to the muon propagation in the rock or to the MACRO angular resolution, 
which are small with respect to BATSE angular resolution.

We also consider possible time correlations between MACRO and BATSE
events.
For the temporal coincidences we use 
both the position information and the time information. 
In order to calculate the background 
we add 200 temporal shifts to the time difference between
the event detected by other experiments and the $\nu$ event in MACRO
considering various time intervals
(the minimum interval is [-4000 s, 4000 s], the maximum
interval is [-80000 s, 80000 s]). We consider time windows
of $\pm 400$ s every 20 s.

We find one event
after 39.4 s from the 4B950922 $\gamma$-ray burst of 22 Sep. 1995
at an angular distance of 17.6$^{\circ}$
and another very horizontal event in coincidence with the
4B940527 $\gamma$-ray burst of 27 May 1994 inside 280 s at $14.9^{\circ}$.
The $90\%$ c.l. muon flux limit is calculated for
a search cone of $10^{\circ}$ around the gamma burst
direction and in an arbitrary time window of $\pm 200$ s.
The choice of this time window is arbitrary because one does not know
{\it a priori} what the duration of the neutrino emission is. Models
of GRB emitters are not yet clear in predicting 
when and for how long neutrinos are
emitted. This is in fact the reason why we have considered 
even a directional analysis of GRBs using no time
information (see previous section).
On the other hand, in this section we are using the time information and
our choice of the time window where we set the upper limit is only
motivated by the fact that this window is larger than the
duration of 97.5$\%$ of the 3B Catalogue GRBs  
(for which the measured durations are available). 
In the chosen search window we find no events 
to be compared to 0.04 expected background events. 
Fig.~\ref{fig15} shows the difference in time between
the detection of an upward-going muon and a GRB
as a function of the cosine of their angular separation.
Two scales are shown: the upper plot is an expanded scale of the lower one.

The corresponding flux upper
limit (90$\%$ c.l.) is 
$0.79 \times 10^{-9}$ cm$^{-2}$ 
upward-going muons per average burst. The limit is almost eight orders of
magnitude lower than the
flux coming from an ``extreme'' topological defect model reported in 
\cite{Halzen}, while according to a model 
in the context of the fireball scenario \cite{Bahcall}
a burst at a distance of 100 Mpc producing 
$0.4 \times 10^{51}$ erg in neutrinos of about $10^{14}$ eV
would produce $\sim 6 \times 10^{-11}$ cm$^{-2}$ upward going muons.

The same analysis on space and time correlations has been performed for 32
{\it Beppo}SAX events; the result is compatible with the atmospheric
neutrino background.

\section{Conclusions}

We have investigated the possibility that the sample of 1100
upward-going muons detected by MACRO since 1989 
shows evidence of a possible neutrino astrophysics source.
We do not find any significant signal with respect to the
statistical fluctuations  of the background due to atmospheric neutrinos
from any of the event directions or from any candidate sources.
We also used the time information to look for correlations with
gamma-ray bursts detected by BATSE and {\it Beppo}SAX.
Having found no excess of events with respect to the
expected background we set muon and neutrino flux upper limits for
point-like sources and for the cumulative search for catalogues 
of sources. These limits have been calculated taking into account
the response of MACRO to various neutrino fluxes from candidate
sources until energies $\gtrsim 100$ TeV.
These limits are for almost all of the considered sources 
the most stringent ones compared to other current
experiments.
They are about 1 order of magnitude higher than values quoted by 
most plausible   
neutrino source models except for the model in \cite{Gondolo}
which is seriously constrained.

{\bf Acknowledgements}\\
We gratefully acknowledge the support of the director and of the staff of the 
Laboratori Nazionali del Gran Sasso and the invaluable assistance of the 
technical staff of the Institutions participating in the experiment. We thank 
the Istituto Nazionale di Fisica Nucleare (INFN), the U.S. Department of 
Energy and the U.S. National Science Foundation for their generous support of 
the MACRO experiment. We thank INFN, ICTP (Trieste) and 
NATO for providing fellowships and grants for non Italian citizens.



\newpage
\begin{table}
\begin{center}
\caption{\label{tab1} Values of the probability 
(given in eq.~\protect\ref{eq:p}) for a neutrino or antineutrino
producing a muon with energy larger than the energy threshold of 1 GeV
as a function of the
neutrino energy.}
\begin{tabular}{ccc}
\tableline\tableline
$E_{\nu}$ (GeV)&$P_{\nu \rightarrow \mu^{-}}$  &
$P_{\bar{\nu} \rightarrow \mu^{+}}$\\ 
\tableline
10     & $8.15 \times 10^{-11}$ & $4.88 \times 10^{-11}$\\
10$^{2}$ & $9.05 \times 10^{-9}$ & $5.87 \times 10^{-9}$\\
10$^{3}$ & $5.79 \times 10^{-7}$ & $3.86 \times 10^{-7}$\\
10$^{4}$ & $1.52 \times 10^{-5}$ & $1.09 \times 10^{-5}$\\
10$^{5}$ & $1.35 \times 10^{-4}$ & $1.17 \times 10^{-4}$\\
\tableline
\end{tabular}
\end{center}
\end{table}

\begin{table}
\begin{center}
\caption{\label{tab2} Fraction of events inside a cone of $3^{\circ}$ half 
width for 2 spectral indices and for 5 zenith angles.}
\begin{tabular}{ccc}
\tableline\tableline
$\cos\theta$ & $\gamma = 2$ & $\gamma = 2.2$ \\ 
\tableline
0.15 & 0.77 & 0.72 \\
0.35 & 0.90 & 0.85 \\
0.55 & 0.91 & 0.87 \\
0.75 & 0.91 & 0.87 \\
0.95 & 0.91 & 0.87 \\ 
\tableline
\end{tabular}
\end{center}
\end{table}

\begin{deluxetable}{ccccccc}
\tablewidth{0pt}
\tablecaption{\label{tab3} $90\%$ c.l. neutrino induced $\mu$-flux limits 
for the MACRO list of 42 sources. Corresponding limits on the
neutrino flux are given in the last column. These limits
are calculated for $\gamma=2.1$ and 
for $E_{\mu} > 1$ GeV including the decrease in efficiency
at very high energies. The reduction factors for a $3^{\circ}$ half width
cone are included. These limits include the effect of 
absorption of neutrinos in the Earth.
The flux upper limits are calculated with the unified approach of 
Feldman \& Cousins \protect\cite{Feldman98}.
B indicates the results of Baksan \protect\cite{Baksan}; 
I the results of IMB \protect\cite{IMB}.}
\tablehead{
\colhead{Source}           & \colhead{$\delta$}      &
\colhead{Events}          & \colhead{Backg.}  &
\colhead{$\nu$ induced}          & \colhead{Previous}    &
\colhead{$\nu$-Flux} \\  
    & \colhead{(degrees)} &
\colhead{in $3^{o}$}  & \colhead{in $3^{o}$}  &
\colhead{$\mu$-Flux}          & \colhead{best}    &
\colhead{limits} \\ 
    &  &  & & Limits& \colhead{$\mu$ limits}   & \\
    &  &  &  & \colhead{$(10^{-14}$ cm$^{-2}$ s$^{-1}$)}  
& \colhead{$(10^{-14}$ cm$^{-2}$ s$^{-1}$)}    &
\colhead{$(10^{-6}$ cm$^{-2}$ s$^{-1}$}) } 
\startdata
 SMC X-1   & -73.5 & 3 & 2.1 & 
0.62 & - & 1.18 \\
 LMCX-2    & -72.0 & 0 & 2.0 & 
0.15 & - & 0.33 \\
 LMCX-4    & -69.5 & 0 & 2.0 
& 0.15 & 0.36 B& 0.29\\
 SN1987A   & -69.3 & 0 & 2.0 
& 0.15 & 1.15 B& 0.31\\
 GX301-2   & -62.7 & 2 & 1.8 
& 0.53 &  -    & 1.10\\
 Cen X-5   & -62.2 & 2 & 1.7 
& 0.55 &  -    & 1.04\\
 GX304-1   & -61.6 & 2 & 1.7 
& 0.54 &  -    & 1.05\\
 CENXR-3   & -60.6 & 1 & 1.7 
& 0.36 & 0.98 I& 0.68\\
 CirXR-1   & -57.1 & 5 & 1.7 
& 1.18 &  -   & 2.21\\
 2U1637-53 & -53.4 & 0 & 1.7 
& 0.19 & -    & 0.36\\
 MX1608-53 & -52.4 & 0 & 1.7 
& 0.20 & -    & 0.38\\
 GX339-4   & -48.8 & 6 & 1.7 
& 1.62 &  -   & 3.00 \\
 ARA XR1   & -45.6 & 3 & 1.6 
& 1.00 &  -   & 1.87\\ 
 VelaP     & -45.2 & 1 & 1.5 
& 0.51 & 0.78 I & 0.94\\
 GX346-7   & -44.5 & 0 & 1.5 
& 0.23 & -     & 0.43\\
 SN1006    & -41.7 & 1 & 1.3 
& 0.56 & - & 1.04\\
 VelaXR-1  & -40.5 & 0 & 1.3 
& 0.26 & 0.45 B& 0.55\\
 2U1700-37 & -37.8 & 1 & 1.3 
& 0.58 & -     & 1.08\\
 L10       & -37.0 & 2 & 1.1 
& 0.91 &  -    & 1.72\\ 
 SGR XR-4  & -30.4 & 0 & 0.9 
& 0.34 &  -    & 0.63\\
 Gal Cen   & -28.9 & 0 & 0.9 
& 0.34 & 0.95 B& 0.65\\ 
 GX1+4     & -24.7 & 0 & 0.9 
& 0.36 &   -   & 0.67\\
 Kep1604   & -21.5 & 2 & 0.9 
& 1.12 & - & 2.12\\ 
 GX9+9     & -17.0 & 0 & 0.9 
& 0.40 & - & 0.75\\
 Sco XR-1  & -15.6 & 1 & 0.9 
& 0.85 & 1.5 B & 1.59\\ 
 Aquarius  & -1.0  & 4 & 0.8 
& 2.48 & - & 4.66\\
 4U0336+01 &  0.6  & 1 & 0.8 
& 1.17 &  -  & 2.19\\
 AQL XR-1  &  0.6  & 0 & 0.8 
& 0.57 & -  & 1.18\\
 2U1907+2  &  1.3  & 0 & 0.8 
& 0.58 & -  & 1.27\\
 SER XR-1  &  5.0  & 0 & 0.7 
& 0.67 & -  & 1.41\\
 SS433     &  5.7  & 0 & 0.7 
& 0.67 & 1.8 B & 1.27\\
 2U0613+09 &  9.1  & 1 & 0.6 
& 1.52 &   - & 3.02\\
 Geminga   &  18.3 & 0 & 0.5 
& 1.12 & 3.1 I & 2.10\\
 Crab      &  22.0 & 1 & 0.4 
& 2.52 & 2.6 B& 4.70\\
 2U0352+30 &  31.0 & 2 & 0.3 
& 5.98 &   -  &11.43\\  
 Cyg XR-1  &  35.2 & 0 & 0.2 
& 3.24 &   -  & 6.24\\
 Her X-1   &  35.4 & 0 & 0.2 
& 3.30 & 4.3 I& 6.96\\
 Cyg XR-2  &  38.3 & 0 & 0.1 
& 4.99 &  -   &10.61\\
 MRK 421   &  38.4 & 0 & 0.1 
& 5.00 & 3.3 I& 9.56\\
 MKN 501   &  40.3 & 0 & 0.1 
& 5.73 & - & 10.69 \\
 Cyg X-3   &  40.9 & 0 & 0.1 
& 6.59 & 4.1 I &  12.49  \\
 Per XR-1  &  41.5 & 0 & 0.1 
& 7.51 & -     & 13.99\\
\enddata
\end{deluxetable}

\clearpage

\begin{table}
\begin{center}
\caption{\label{tab4} Percentage increase (+) or decrease (-) of
MACRO exposure between a source producing
a neutrino spectrum with $\gamma=$ 2.3, 2.5, 2.7 (such in the case of
atmospheric neutrinos with $E_{\nu} \lesssim 10$ GeV) and 3.7 
(such in the case of high energy atmospheric neutrinos) 
with respect to the  $\gamma=2.1$ spectrum.}
\begin{tabular}{ccccc}
\tableline\tableline
Declination & Difference  (\%) &  Difference (\%)  & Difference  (\%) 
& Difference (\%)\\ 
            & $\gamma=2.3$ & $\gamma=2.5$& $\gamma=2.7$ & 
$\gamma=3.7$ \\ 
\tableline
$-90^{\circ}$  & $+4.1$ & $+5.9$ & $+5.2$ & $-23.8$ \\
$-80^{\circ}$  & $+2.8$ & $+4.2$ & $+3.9$ & $-18.7$ \\
$-70^{\circ}$  & $+3.0$ & $+4.6$ & $+4.5$ & $-18.7$ \\
$-60^{\circ}$  & $+2.9$ & $+4.5$ & $+4.4$ & $-18.9$ \\
$-50^{\circ}$  & $+2.7$ & $+4.0$ & $+3.6$ & $-19.9$ \\
$-40^{\circ}$  & $+3.0$ & $+4.6$ & $+4.4$ & $-18.6$ \\
$-30^{\circ}$  & $+3.0$ & $+4.5$ & $+4.2$ & $-19.0$ \\
$-20^{\circ}$  & $+3.0$ & $+4.4$ & $+3.9$ & $-20.0$ \\
$-10^{\circ}$  & $+3.3$ & $+4.9$ & $+4.4$ & $-20.4$ \\
  $0^{\circ}$  & $+3.7$ & $+5.6$ & $+5.3$ & $-19.5$ \\
  $10^{\circ}$ & $+3.8$ & $+5.8$ & $+5.7$ & $-19.5$ \\
  $20^{\circ}$ & $+4.1$ & $+6.6$ & $+7.1$ & $-15.8$ \\
  $30^{\circ}$ & $+5.1$ & $+7.8$ & $+7.8$ & $-17.2$ \\
  $40^{\circ}$ & $+5.0$ & $+7.4$ & $+6.9$ & $-16.7$ \\ 
\tableline
\end{tabular}
\end{center}
\end{table}

\begin{table}
\begin{center}
\caption{\label{tab5} Flux upper limits on muon fluxes from neutrino production
for various catalogues at 90$\%$ c.l. for $E_{\mu} > 1$ GeV
and for an assumed spectral index of the neutrino flux of
$\gamma=2.1$. The catalogue, the number of souces in it, the average exposure
which is the denominator in eq.~\ref{eq:newlim},
the muon flux upper limit
and the neutrino flux upper limit are given. Flux upper limits are calculated
according to eq.~\ref{eq:upp1} and eq.~\ref{eq:upp2}.}
\begin{tabular}{ccccc}
\tableline\tableline
Catalog & Num.    & Average Exposure & 
$\nu$-induced $\mu$ flux & $\nu$ flux\\ 
        & sources & (cm$^{2}$ s) &  
limit     & limit\\
        &         &       &   (cm$^{-2}$s$^{-1})$
&(cm$^{-2}$s$^{-1}$)\\ 
\tableline
MACRO list                 & 42  & $4.67 \cdot 10^{14}$& 
$3.06 \cdot 10^{-16}$ & $5.79 \cdot 10^{-8}$\\
SNRs \protect\cite{Green}      & 220 & $2.35 \cdot 10^{14}$& 
$2.63 \cdot 10^{-16}$ & $ 5.00 \cdot 10^{-8}$\\ 
Blazar \protect\cite{Padovani} & 181 & $2.77 \cdot 10^{14}$ &
$5.44 \cdot 10^{-16}$ & $1.03 \cdot 10^{-7}$\\
BATSE \protect\cite{BATSE}     & 2527& $3.37 \cdot 10^{14}$ & 
$1.68 \cdot 10^{-16}$ & $3.17 \cdot 10^{-8}$\\
EGRET \protect\cite{EGRET}     & 271 & $ 3.66 \cdot 10^{14}$ &
$2.03 \cdot 10^{-16}$ & $3.82 \cdot 10^{-8}$\\
{\it Beppo}SAX \protect\cite{Pian} & 32 & $3.41 \cdot 10^{14}$ &
$6.84 \cdot 10^{-16}$ & $1.27 \cdot 10^{-7}$\\
NovaeX \protect\cite{Masetti}  & 29  & $4.97 \cdot 10^{14}$ &
$5.34 \cdot 10^{-16}$ & $1.01 \cdot 10^{-7}$ \\
\tableline
\end{tabular}
\end{center}
\end{table}

\begin{table}
\begin{center}
\caption{\label{tab6} Upper limits ($90\%$c.l.) on upward muon fluxes induced
by neutrinos from annihilation of neutralinos trapped in the Galactic Center
for neutralino masses $m_{\chi} =$ 60, 100, 200, 500 GeV and
various search cones around its direction. These limits
are for $E_{\mu} > 1$ GeV and upper limits are calculated according to 
\protect\cite{Feldman98}. 
These limits constrain strongly the model by Gondolo and Silk
\protect\cite{Gondolo} (compare them to Fig. 3 in their paper.)}
\begin{tabular}{cccccccc}
\tableline\tableline
Cone & Data &  Background & \multicolumn{5}{c}{$m_{\chi}$ (GeV)}\\
     &      &             & 60 & 100 & 200 & 500 & 1000\\
     &      &             & \multicolumn{5}{c}{$\mu$-Flux limits} \\
     &      &             & \multicolumn{5}{c}{$\times 10^{-15}$
cm$^{-2}$s$^{-1}$} \\
\tableline
$3^{\circ}$  & 0  & 0.9  & 3.67 & 3.42 & 3.26 & 3.21 & 3.21 \\
$5^{\circ}$  & 1  & 2.6  & 4.70 & 4.37 & 4.17 & 4.11 & 4.11 \\
$10^{\circ}$ & 10 & 10.3 & 13.66 & 12.71 & 12.11 & 11.96 & 11.94 \\
\tableline
\end{tabular}
\end{center}
\end{table}

\newpage
\epsfig{file=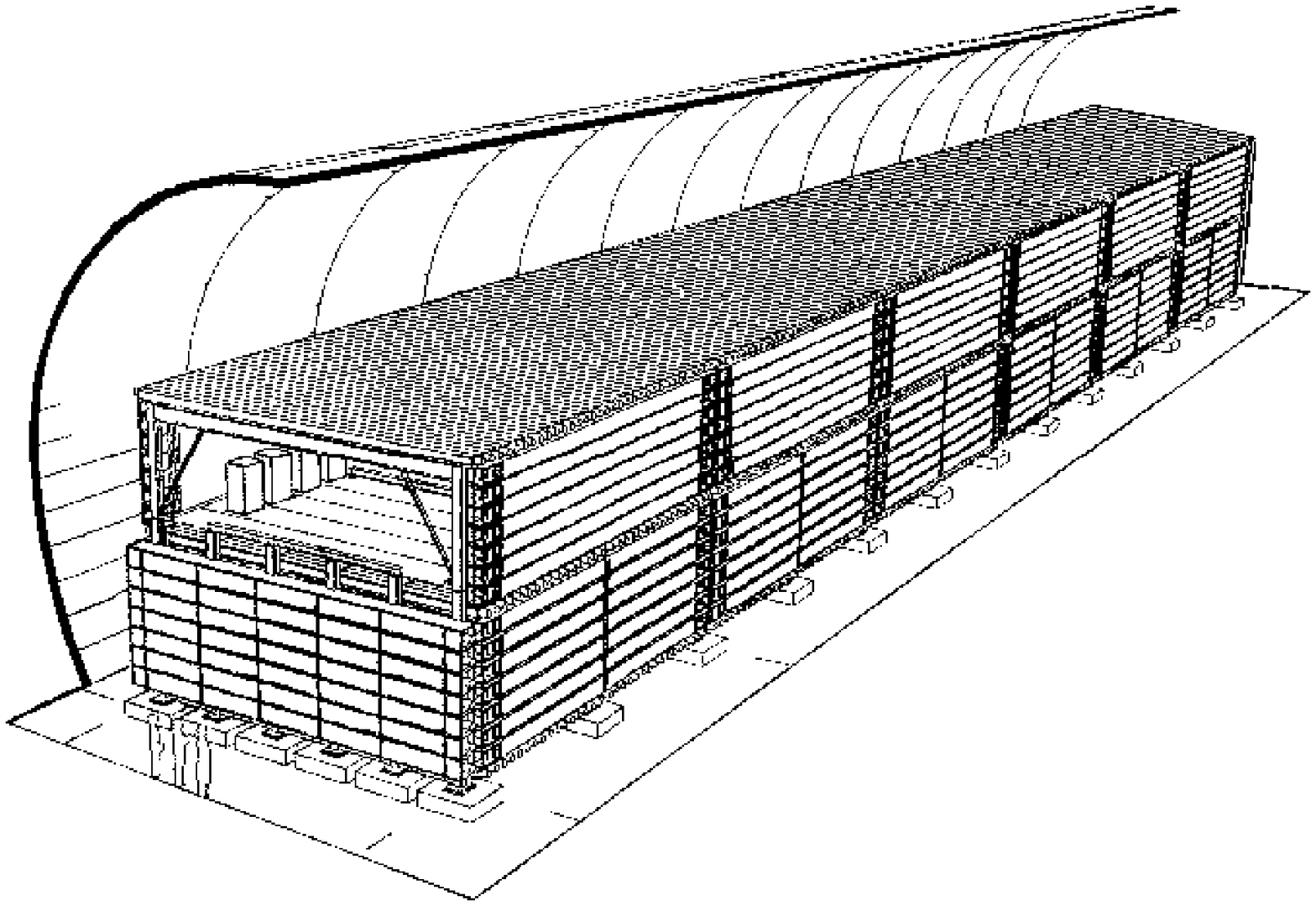,height=9.cm,width=16.cm}
\figcaption{Layout of the MACRO detector. \label{fig1}}

\epsfig{file=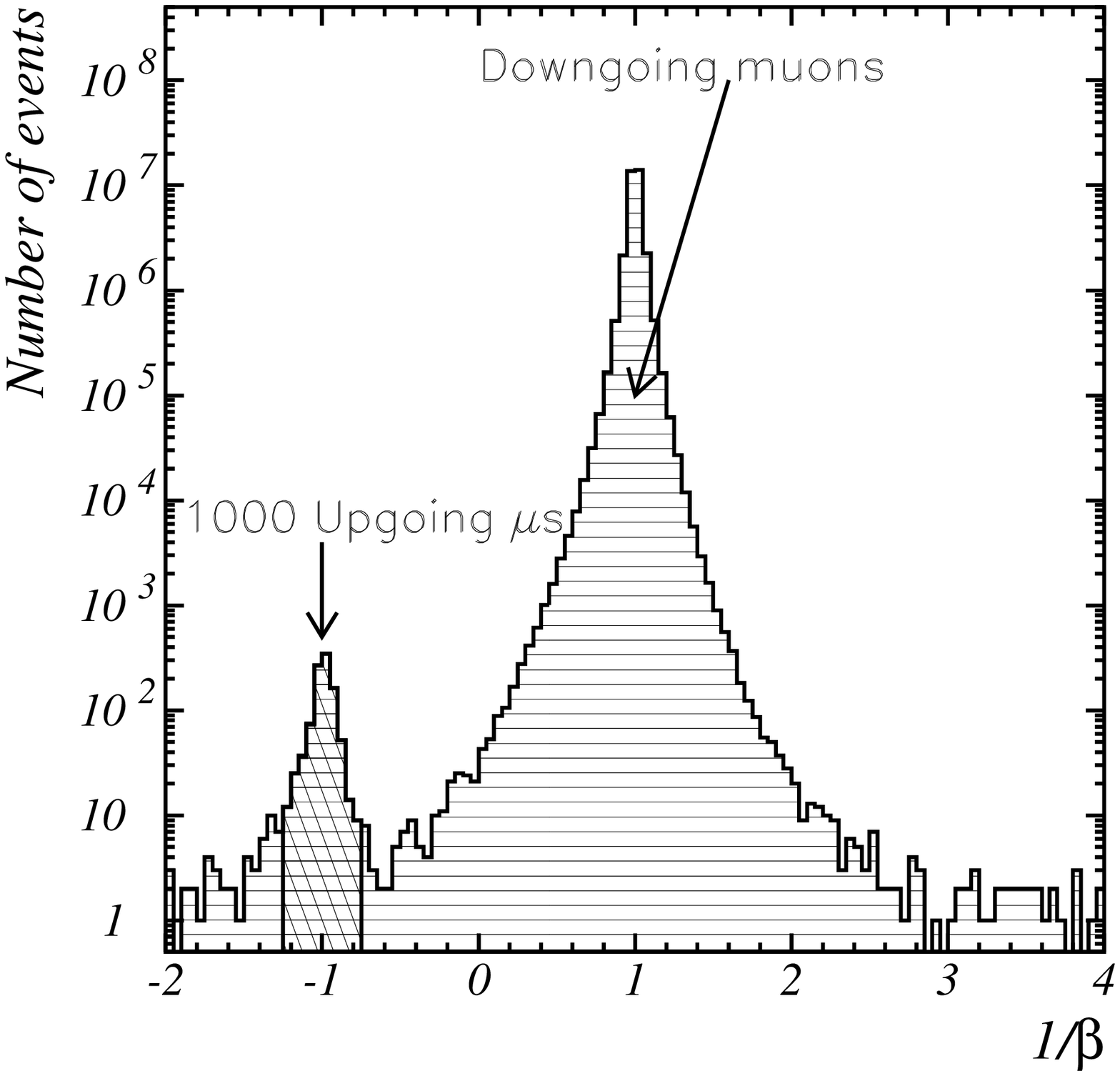,height=9.cm,width=15.cm}
\figcaption{The $1/\beta$ 
distribution for the muon data sample collected with
the full detector; the number of downgoing muons 
is $\sim 33.8 \times 10^{6}$. \label{fig2}}

\epsfig{file=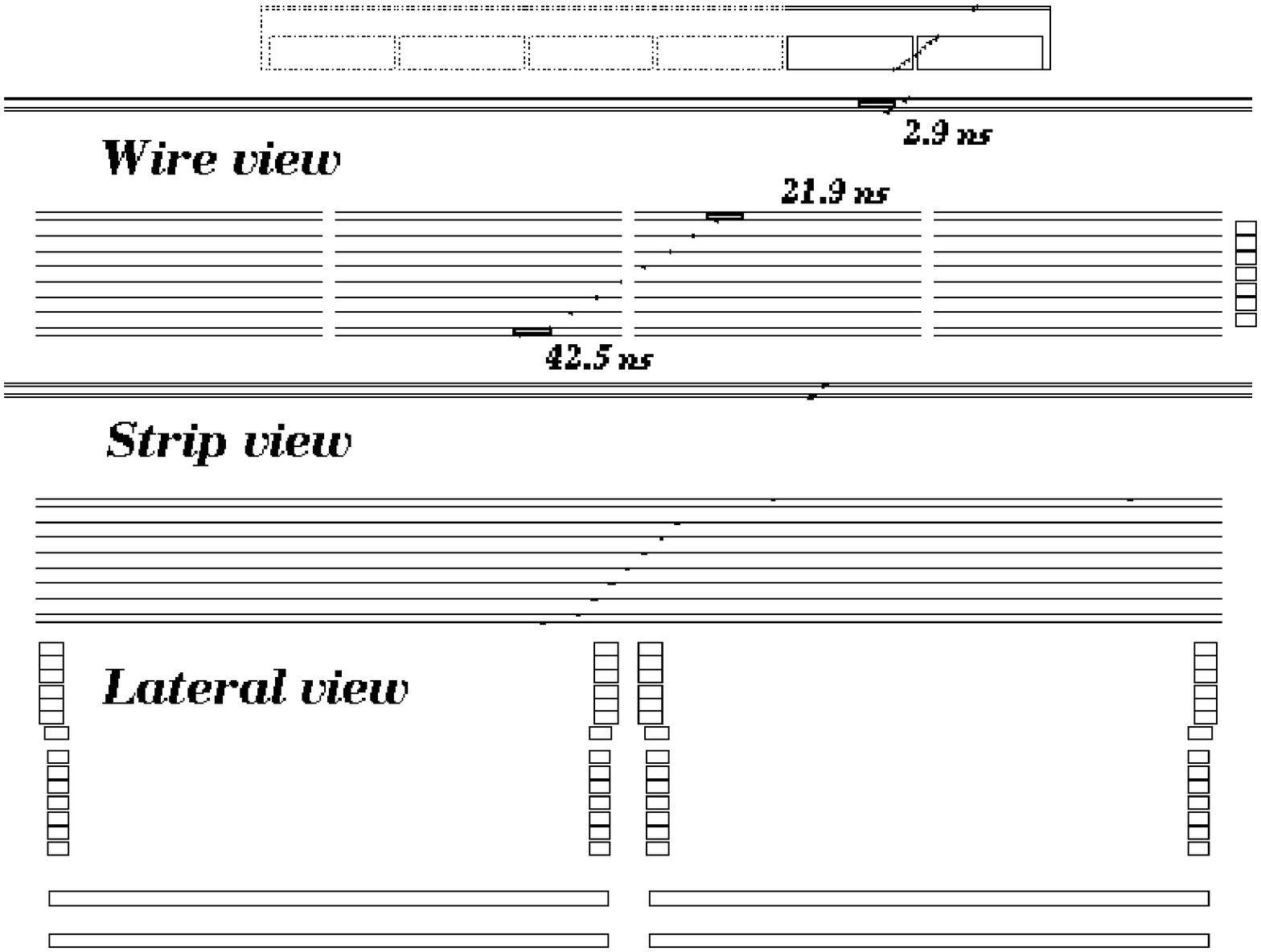,height=16.cm,width=16.cm}
\figcaption{One upward-going muon produced by a neutrino interaction 
in the rock below the MACRO detector. The wire, strip and lateral views are 
shown and the times in nanoseconds are indicated near the scintillator
counters hit by the track. The first hit counter corresponds to the larger 
time. \label{fig3}}

\epsfig{file=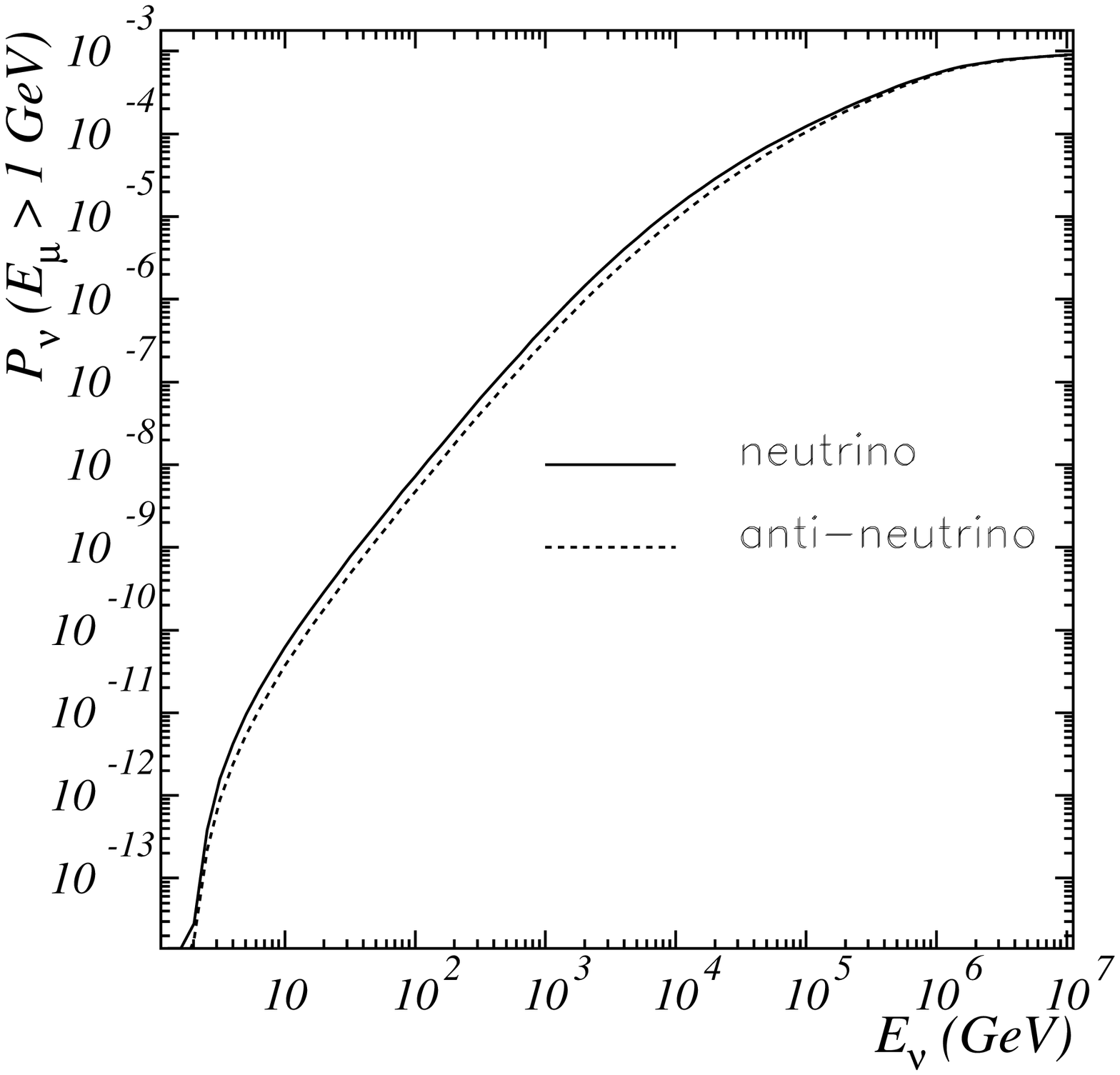,height=8.cm,width=15.cm}
\figcaption{Probability that a $\nu$ with $E_{\nu}$ crossing the
detector produces a muon above threshold. Solid line: neutrino;
dotted line: anti-neutrino. \label{fig4}}      

\epsfig{file=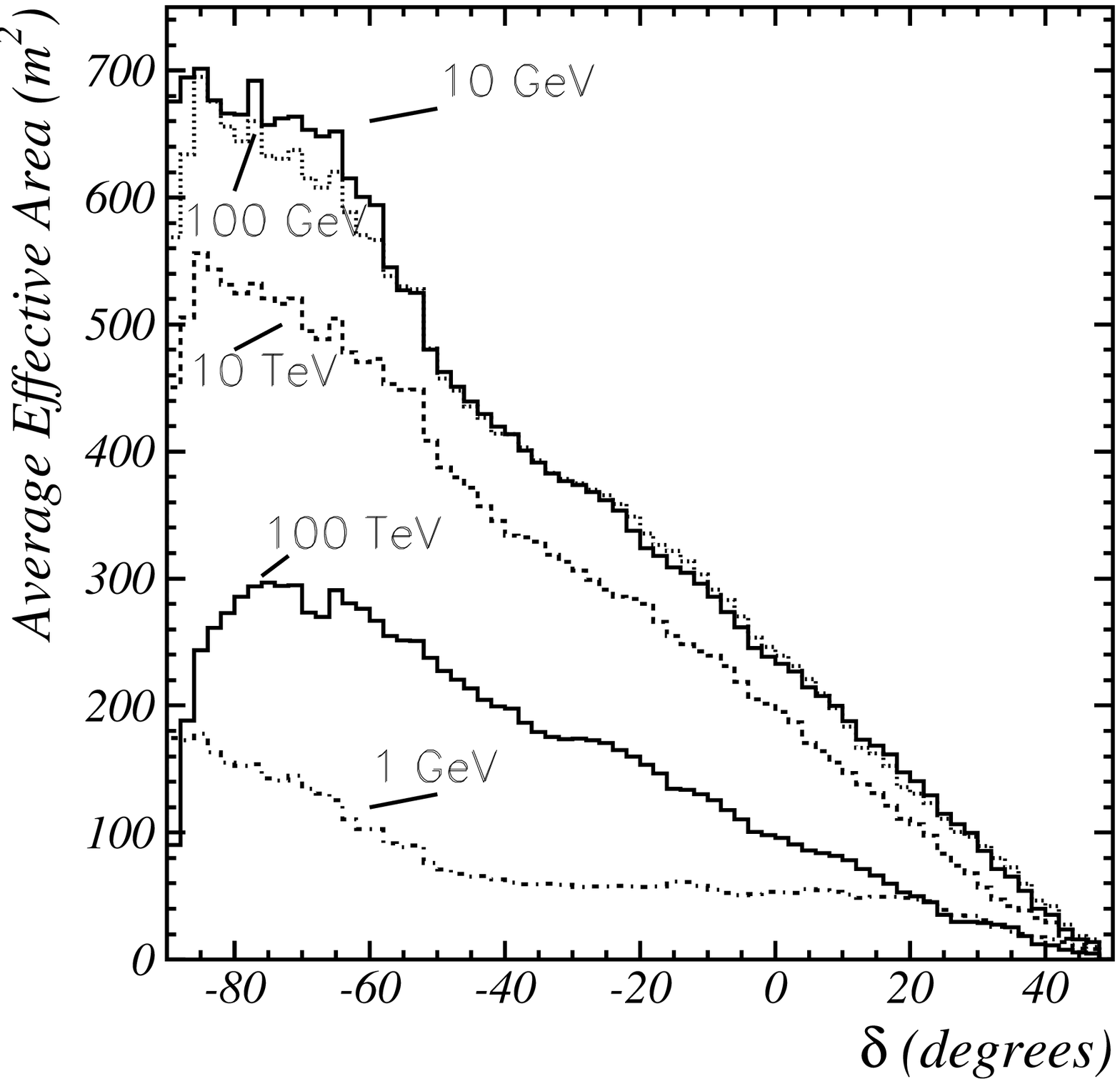,height=9.cm,width=15.cm}
\figcaption{MACRO average effective area as a function of declination 
for various muon energies.
From top to bottom lines: 10 GeV (solid line), 100 GeV (dotted 
line), 10 TeV (dashed line), 100 TeV (solid line) and 1 GeV 
(dot-dashed line). \label{fig5}}

\epsfig{file=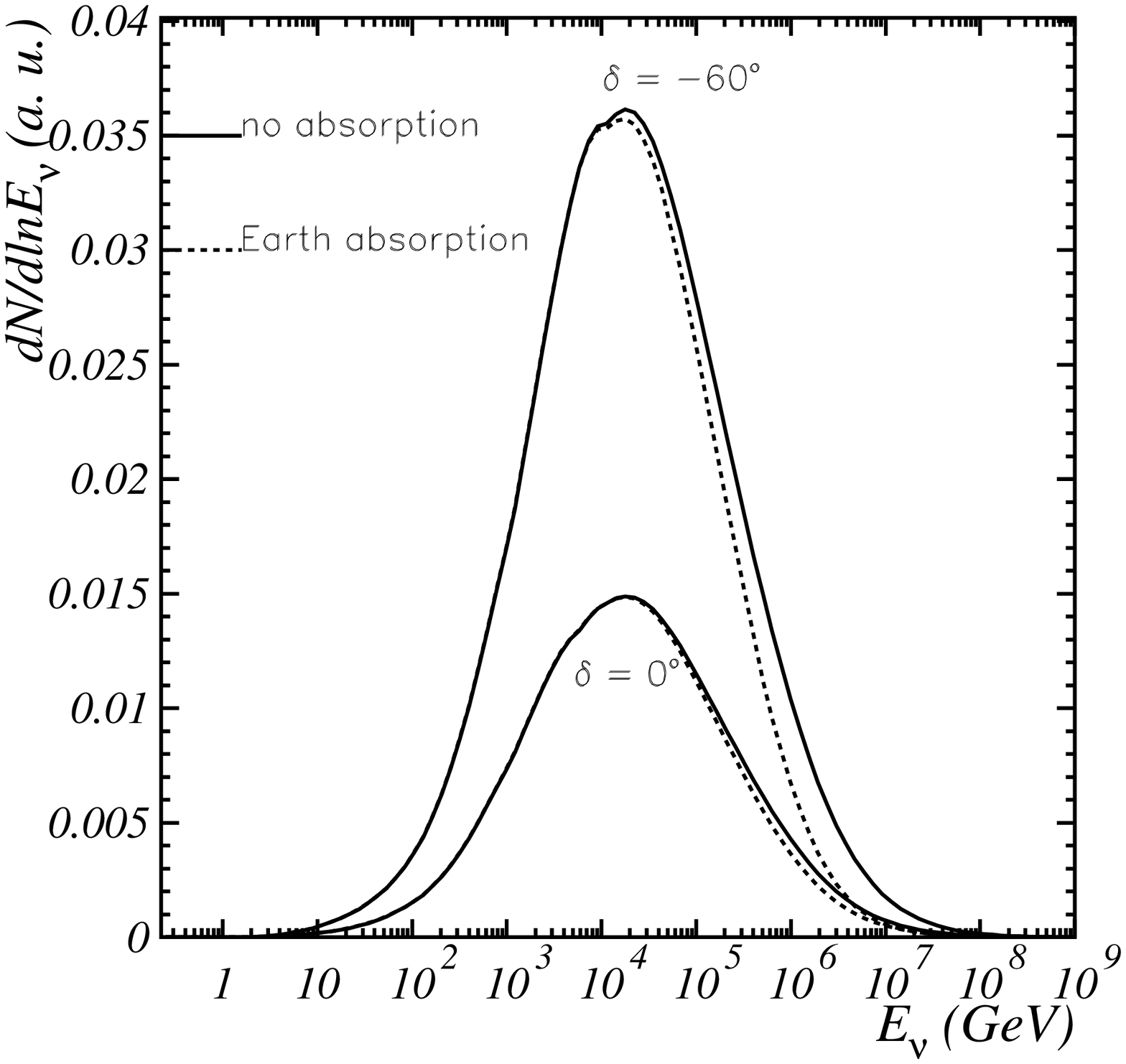,height=8.5cm,width=15.cm}
\figcaption{MACRO response curve, 
which is the differential rate of neutrinos which induce detectable muons
as a function of the neutrino energy, 
for 2 sources at declinations 
$-60^{\circ}$ and $0^{\circ}$. Solid (dotted) lines do not (do) 
include Earth absorption. The normalization of the neutrino fluxes 
is arbitrary. \label{fig6}}

\epsfig{file=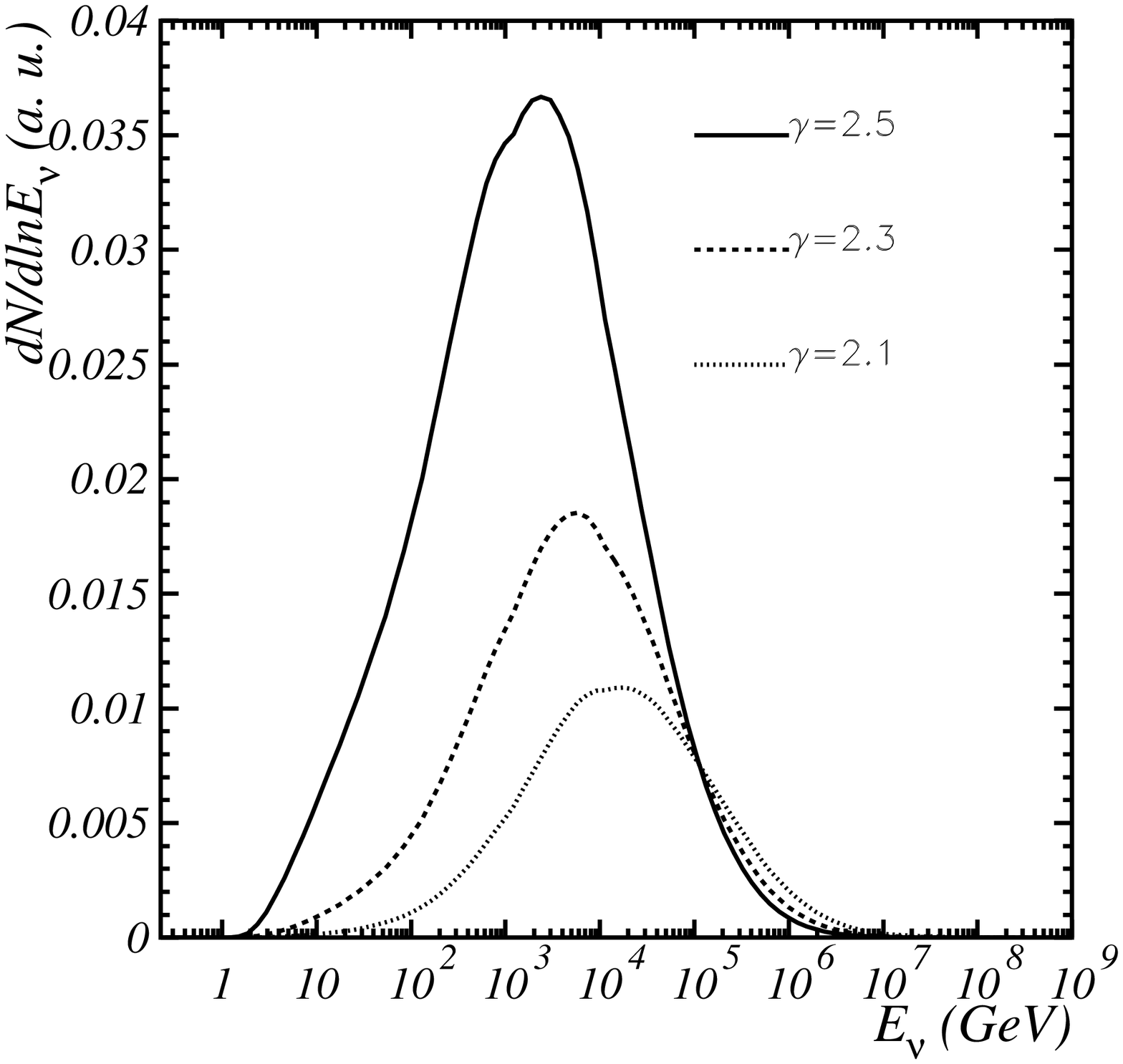,height=9.cm,width=15.cm}
\figcaption{MACRO response curves
for spectral indices ($\gamma $=2.1, 2.3, 2.5)
for a source at declination $-60^{\circ}$. Earth absorption is included.
\label{fig7}}

\epsfig{file=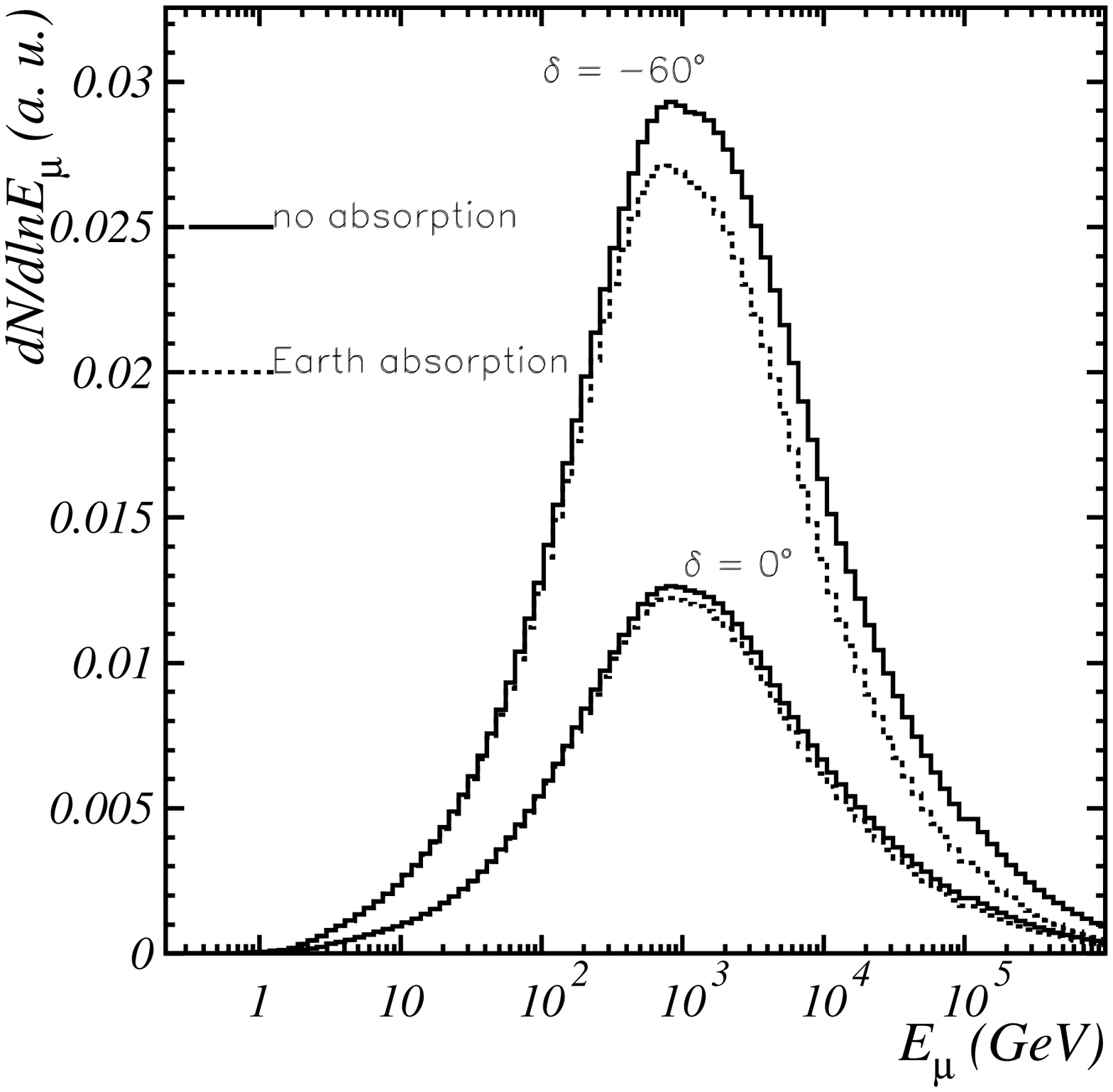,height=8.5cm,width=15.cm}
\figcaption{Differential rate of detected muons 
as a function of muon energy in
MACRO for spectral indices $\gamma $=2.1 and for
sources at declinations $-60^{\circ}$ and $0^{\circ}$
with (dotted line) and without (solid line) absorption. \label{fig8}}

\epsfig{file=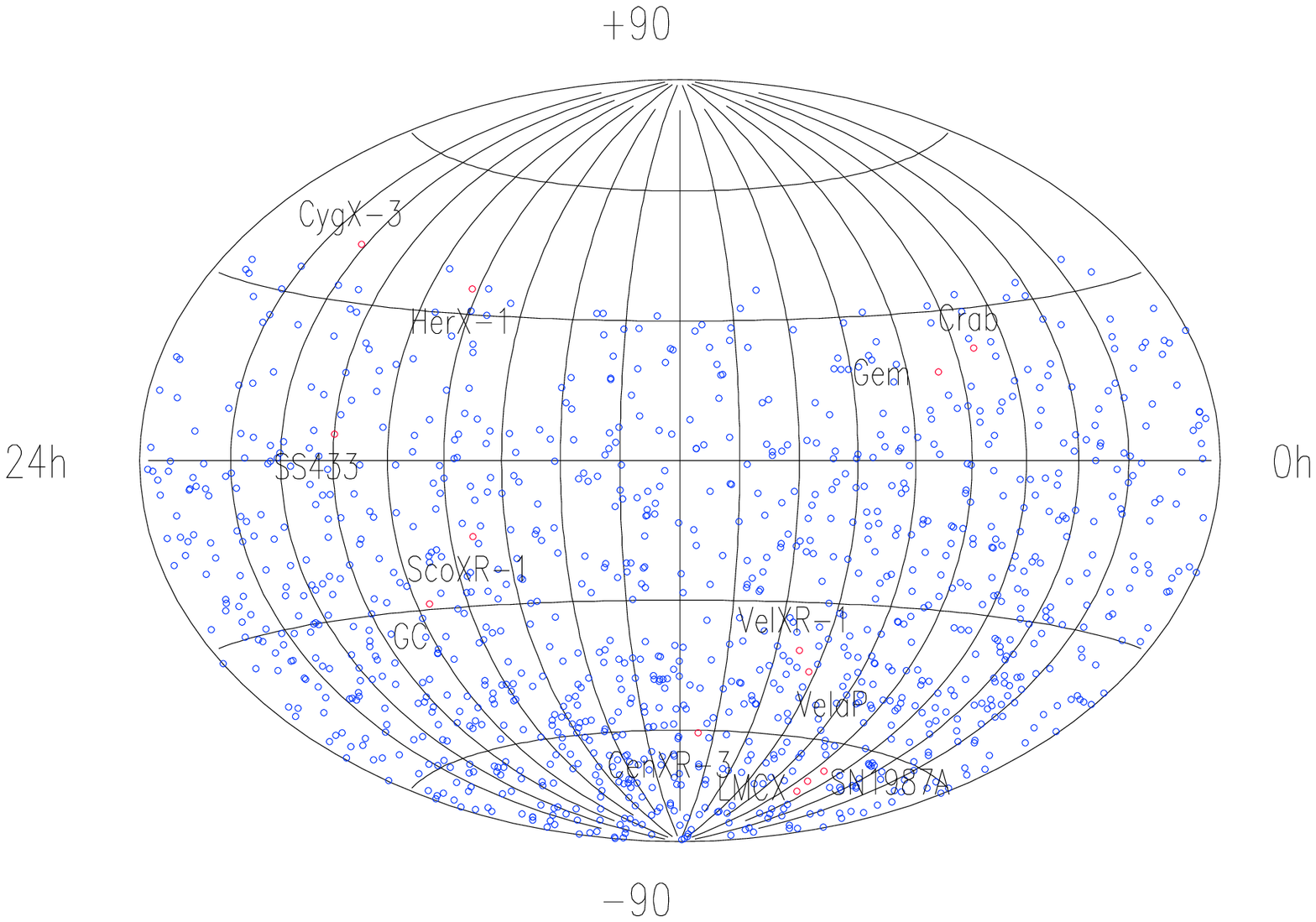,height=9.cm,width=15.cm}
\figcaption{Upward-going muon distribution in equatorial coordinates
(1100 events).  
\label{fig9}}

\epsfig{file=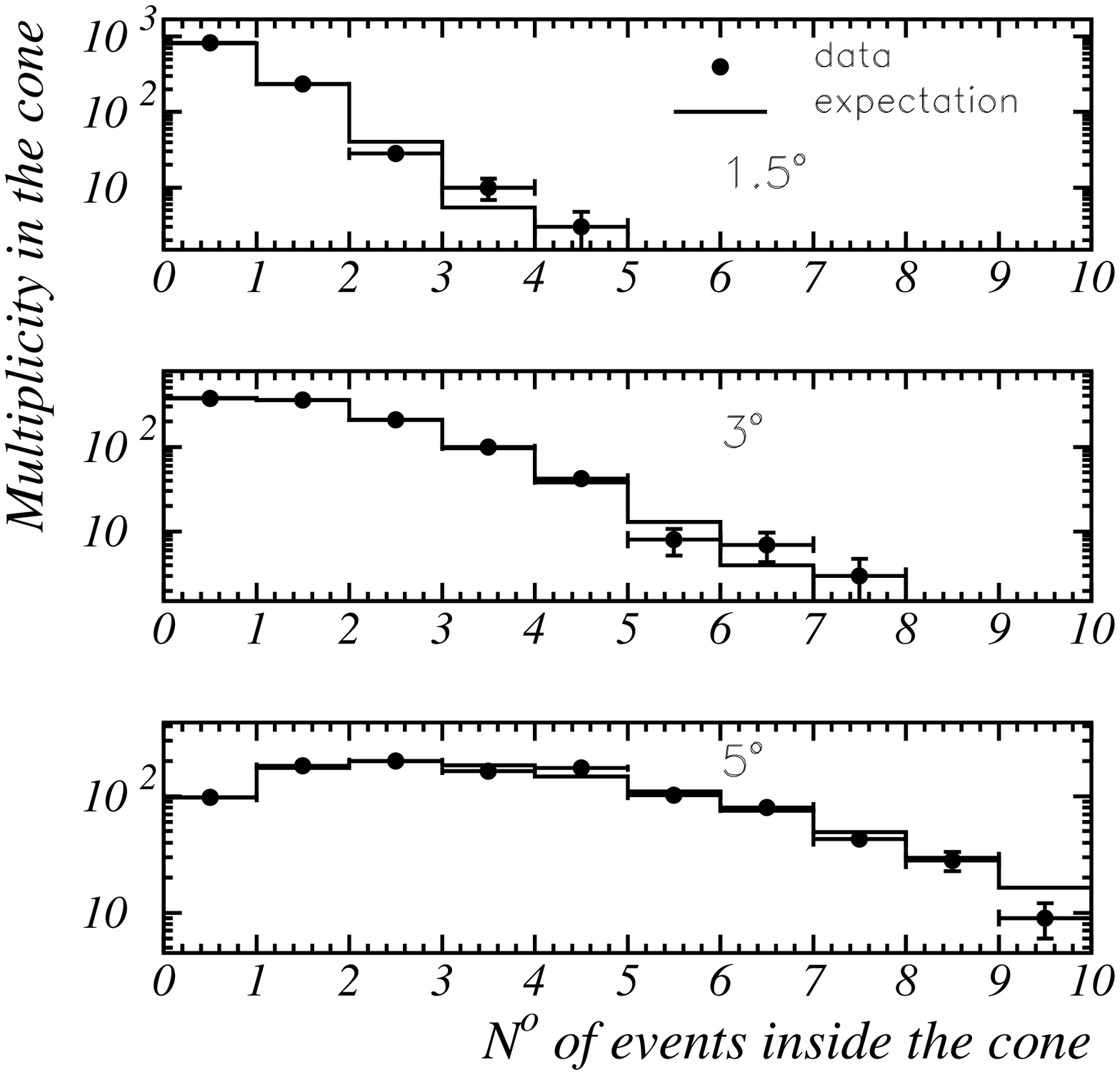,height=13.cm,width=15.cm,}
\figcaption{On the x axis there is the number of events falling in cones 
of half width $1.5^{\circ}$, $3^{\circ}$, $5^{\circ}$ (from top to bottom 
plot) around the direction of any muon.
The y axis depends on the total number of events considered.
Black circles: data. Solid line: simulation. \label{fig10}}

\epsfig{file=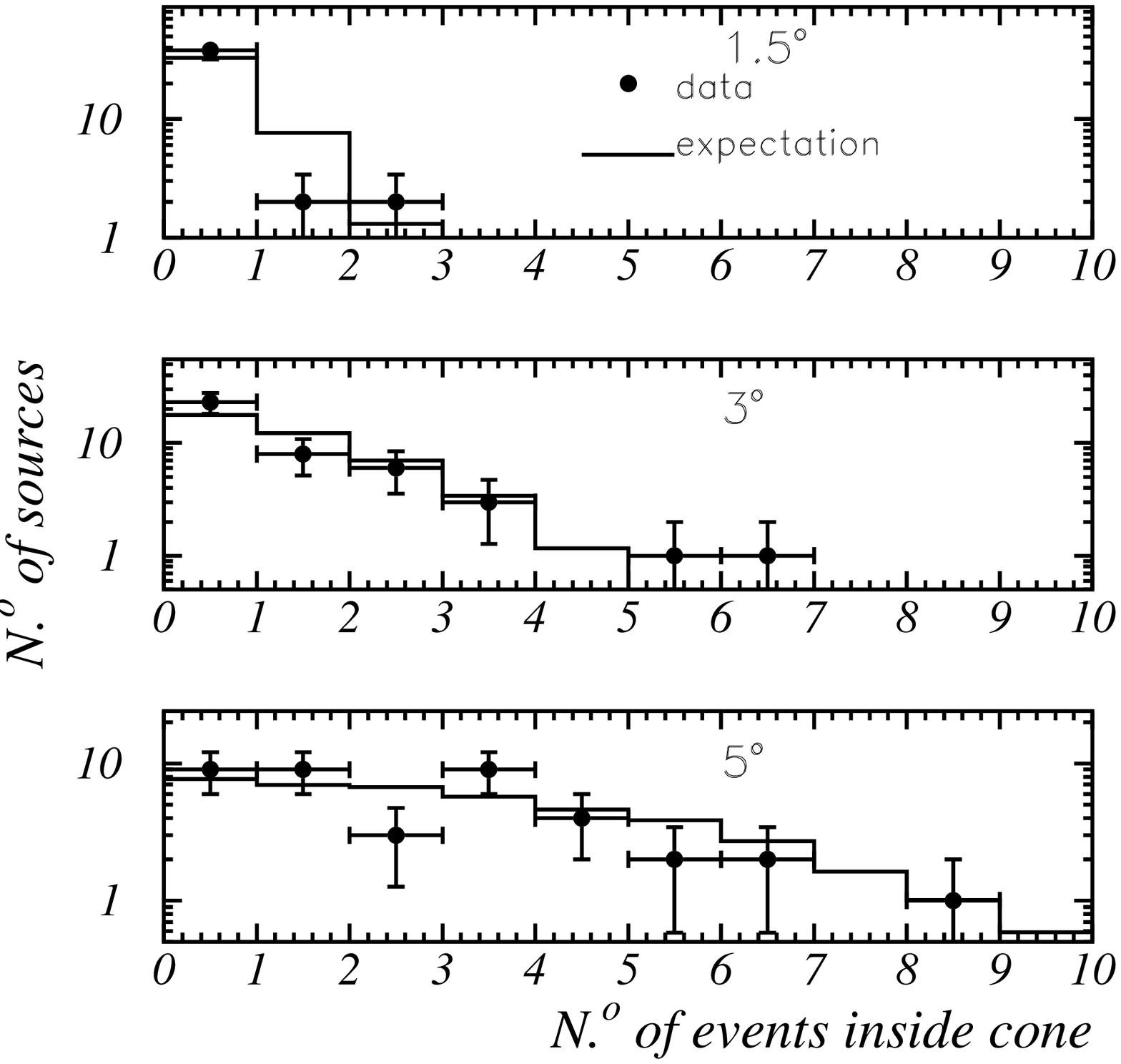,height=13.cm,width=15.cm,}
\figcaption{On the x axis there is the number of events falling in cones 
of half width $1.5^{\circ}$, $3^{\circ}$, $5^{\circ}$ (from top to bottom) 
around the direction of the 42
sources considered.
The y axis depends on the total number of sources considered.
Black circles: data. Solid line: simulation. \label{fig11}}

\epsfig{file=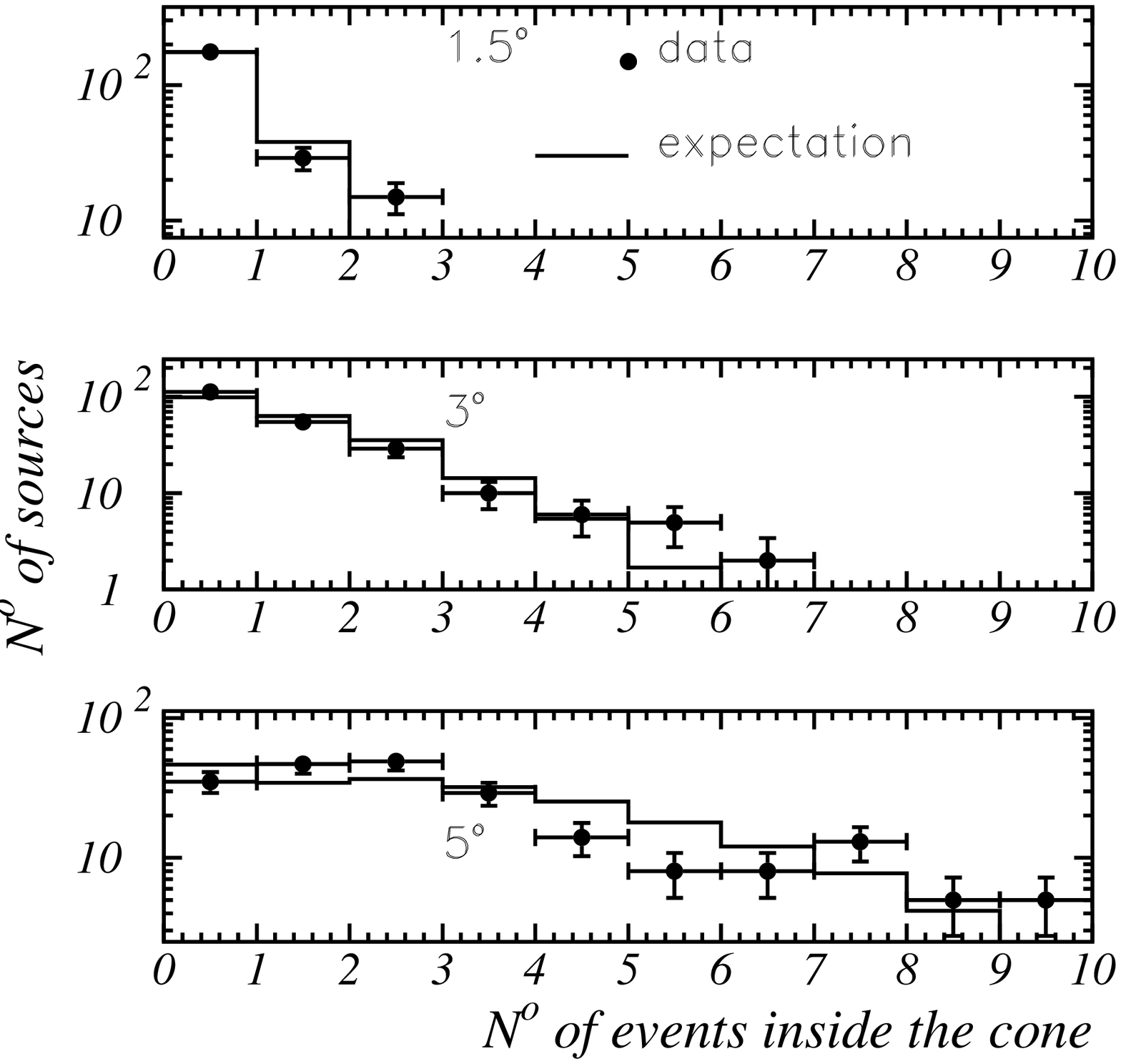,height=13.cm,width=15.cm}
\figcaption{On the x axis there is the number of events falling in cones 
of half width $1.5^{\circ}$, $3^{\circ}$, $5^{\circ}$ (from top to bottom) 
around the direction of 220 SNRs
from \protect\cite{Green}.
The y axis depends on the total number of sources considered.
Black circles: data. Solid line: simulation. \label{fig12}}

\epsfig{file=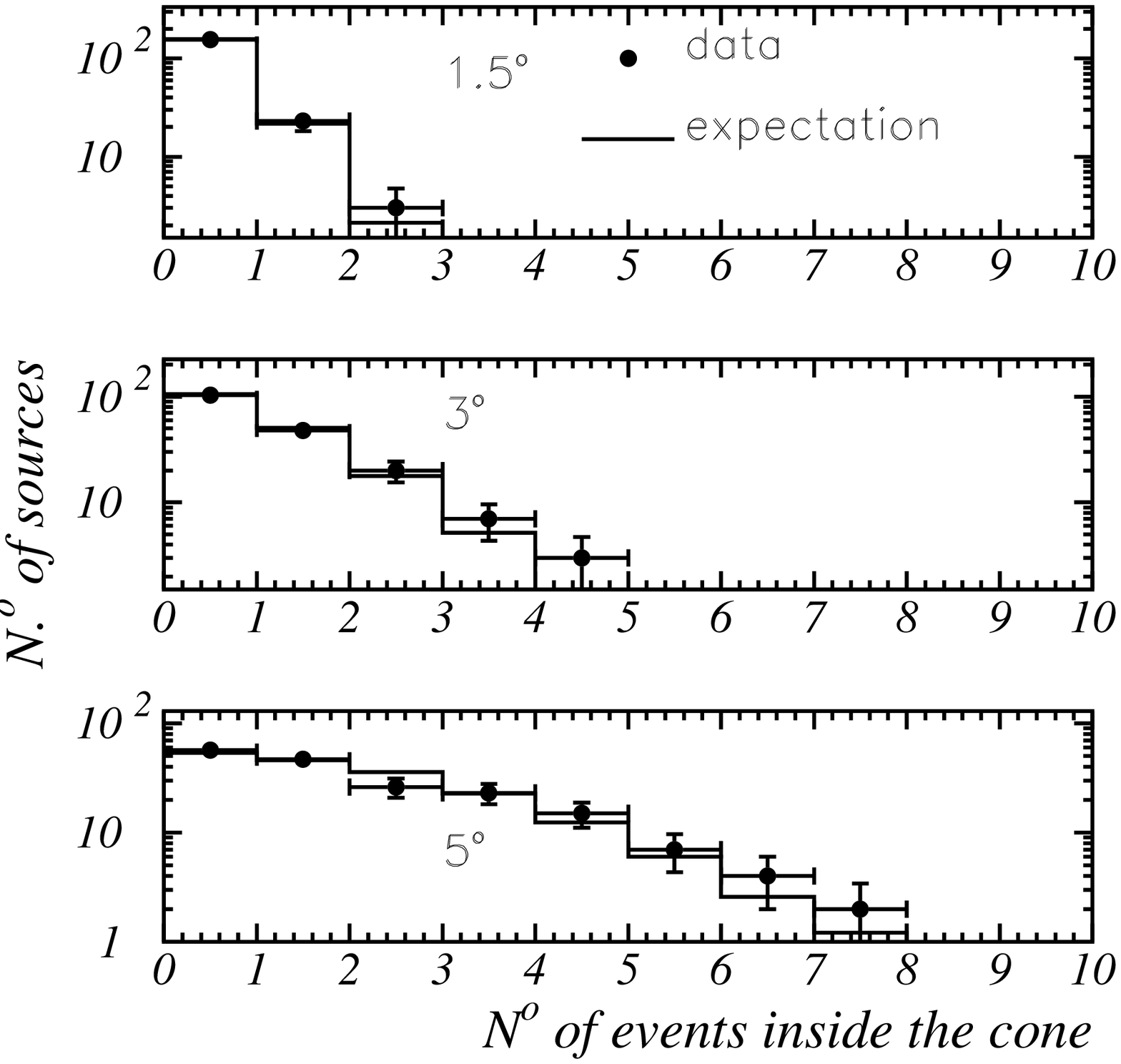,height=13.cm,width=15.cm}
\figcaption{On the x axis there is the number of events falling in cones 
of $1.5^{\circ}$, $3^{\circ}$, $5^{\circ}$ around the direction of 181 
blazars
from \protect\cite{Padovani}.
The y axis depends on the total number of sources considered.
Black circles: data. Solid line: simulation. \label{fig13}}

\epsfig{file=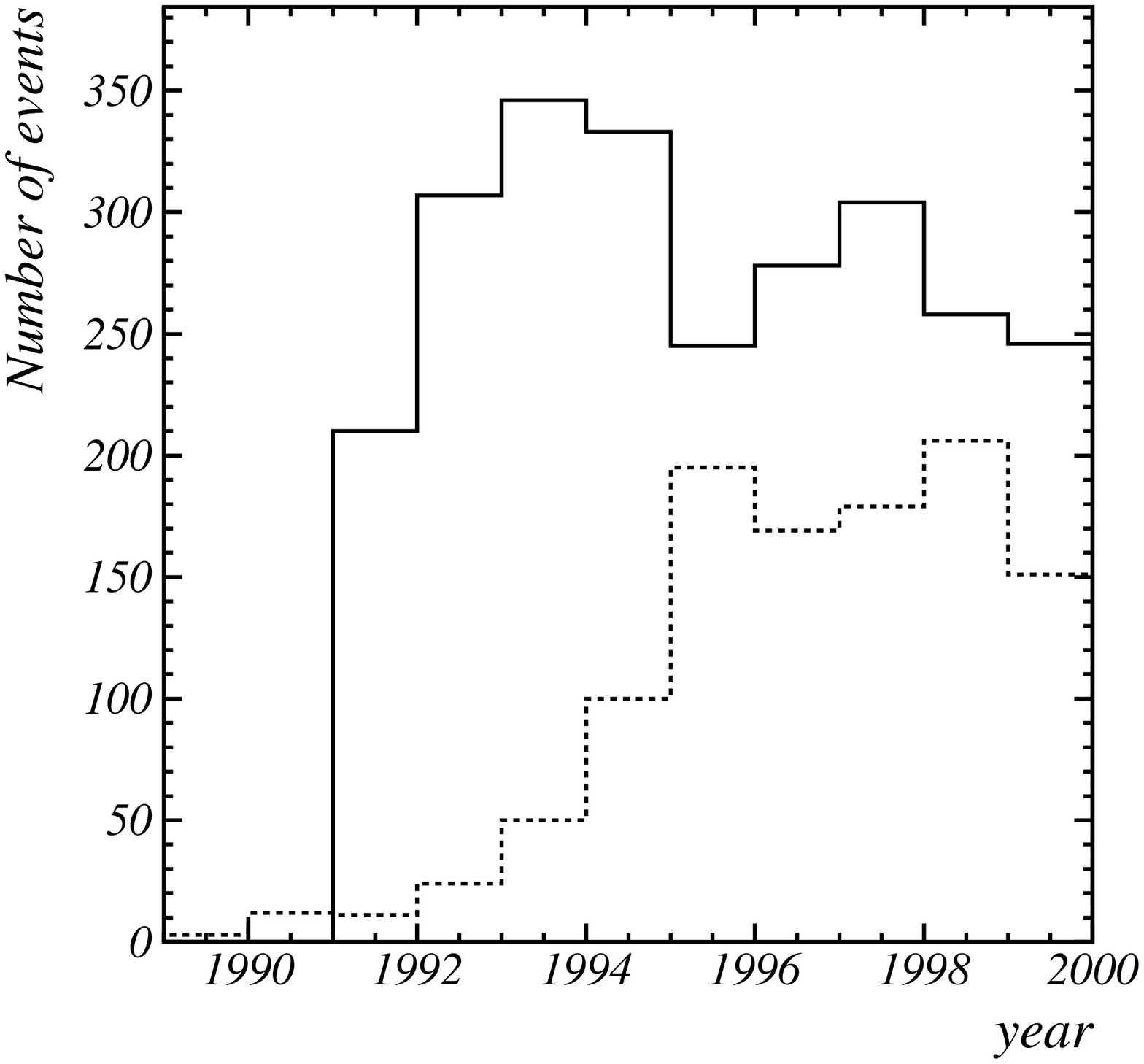,height=9.cm,width=15.cm,}
\figcaption{MACRO events (dashed line) and BATSE GRBs (solid line) 
as a function of the year. \label{fig14}}

\epsfig{file=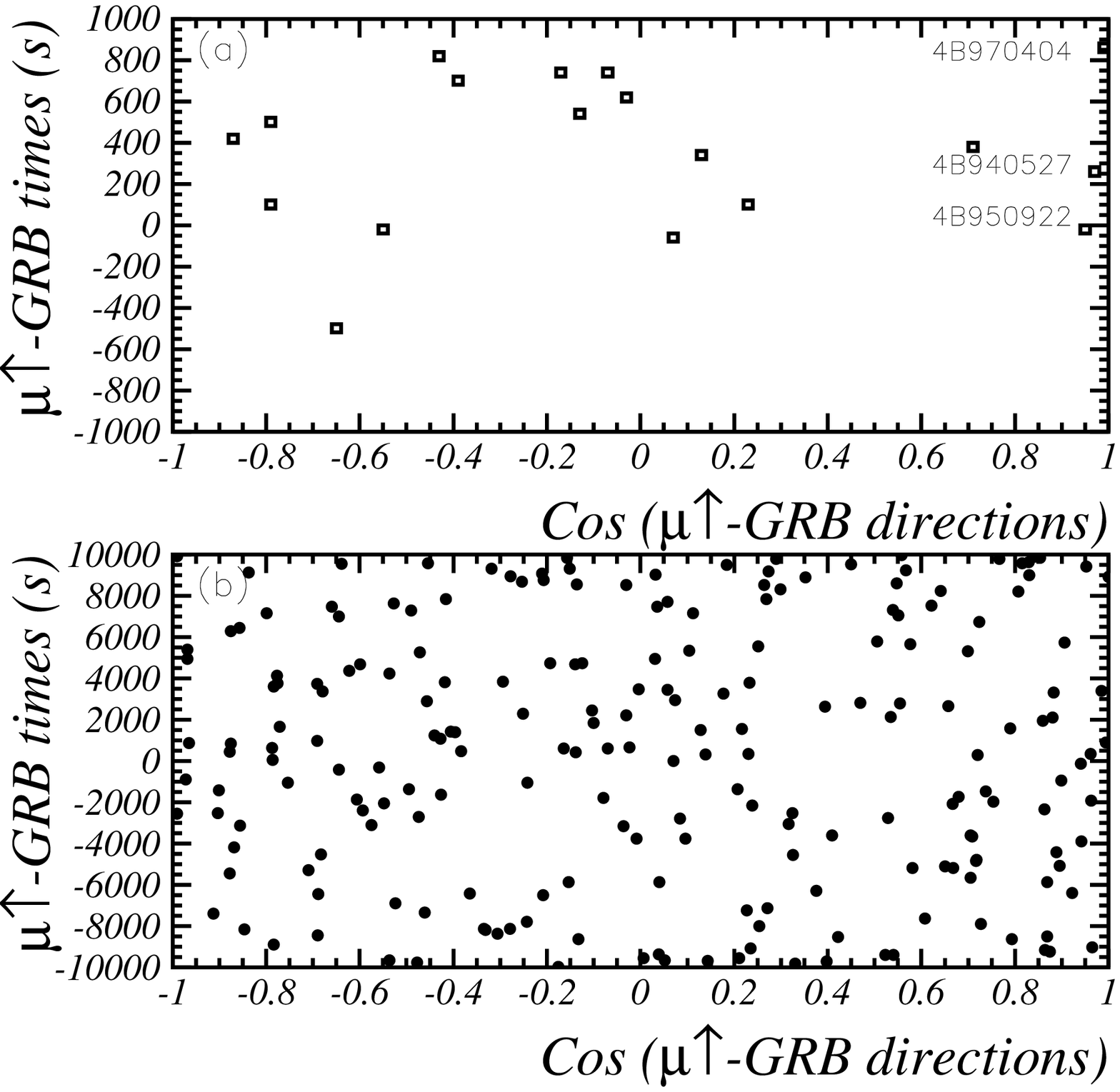,height=13.cm,width=15cm}
\figcaption{Difference in detection times vs the cosine of the
angular separation between MACRO upward-going muon events and 
BATSE gamma ray bursts. (a) and (b) have different time scales.
The BATSE GRB of 22 Sep. 1995 (4B 950922) was detected 39.4 s before
one MACRO event at an angular distance of 17.6$^{\circ}$
and the BATSE (4B 940527) GRB of 27 May 1994 
after 280 s another very horizontal MACRO event at $14.9^{\circ}$. 
In (a) they are indicated 
with the name of the bursts. \label{fig15}}

\end{document}